\documentclass[onefignum,onetabnum]{siamart171218}
\usepackage{amsfonts}
\usepackage{graphicx}
\usepackage{placeins}

\newcommand{\tr}[1]{\mathrm{tr}\,{#1}}
\newcommand{\dv}[1]{\mathrm{div}\,{#1}}
\newcommand{\cl}[1]{\mathrm{curl}\,{#1}}

\newcommand{\dvr}{\mathrm{div}}

\newcommand{\beq}{\begin{equation}}
\newcommand{\eeq}{\end{equation}}

\newsiamremark{remark}{Remark}
\newsiamremark{hypothesis}{Hypothesis}
\crefname{hypothesis}{Hypothesis}{Hypotheses}
\newsiamthm{claim}{Claim}

\headers{Phase Transitions in Nematics}{Golovaty, Kim, Lavrentovich, Novack, Sternberg}

\title{Phase Transitions in Nematics: Textures with Tactoids and Disclinations}

\author{Dmitry Golovaty\thanks{Department of Mathematics, The University of Akron, Akron, OH, USA 
  (\email{dmitry@uakron.edu}).}
\and Young-Ki Kim\thanks{Robert Frederick Smith School of Chemical and Biomolecular Engineering, Cornell University, Ithaca, NY, USA 
  (\email{yk756@cornell.edu}).}
\and Oleg D.~Lavrentovich\thanks{Chemical Physics Interdisciplinary Program, Liquid Crystal Institute, Kent State University, OH, USA 
  (\email{olavrent@kent.edu}).}
\and Michael Novack\thanks{Department of Mathematics, Indiana University, Bloomington, IN, USA 
  (\email{mrnovack@indiana.edu}).}
\and Peter Sternberg\thanks{Department of Mathematics, Indiana University, Bloomington, IN, USA 
  (\email{sternber@indiana.edu}).}}

\begin{document}

\maketitle

\begin{abstract}
We demonstrate that a first order isotropic-to-nematic phase transition in liquid crystals can be succesfully modeled within the generalized Landau-de Gennes theory by selecting an appropriate combination of elastic constants. The numerical simulations of the model established in this paper qualitatively reproduce the experimentally observed configurations that include interfaces and topological defects in the nematic phase. 
\end{abstract}


\section{Introduction}
\label{sec:intro}
Phase transitions between ordered and isotropic states in nematic liquid crystals are of interest both for technological reasons as well as because nematics offer one of the  simplest experimental  systems where interfaces and topological defects may coexist. In particular, a {\em uniaxial} nematic liquid  crystal typically consists of asymmetrically-shaped molecules that under appropriate conditions tend to align roughly in the same direction. The preferred molecular orientation in a given region occupied by a uniaxial nematic is often described in terms of the unit vector field $n$, called the {\em director}. A more precise description takes into account the non-polar character of ordering, e.g., because the probabilities of finding the head or the tail of a molecule in a given direction are equal. This description replaces the director $n$ with a projection matrix $n\otimes n$ which can be identified with an element of the projective space and is invariant with respect to reversal of orientation $n\to-n$.

Two different types of topological defects can be present in a nematic in three dimensions: $0$-dimensional point defects also known as vortices or nematic hedgehogs and $1$-dimensional disclinations. Point defects may also be present on the surface of a nematic; such defects are known as boojums. There is a large body of literature devoted to the study of nematic defects and we refer the reader to \cite{kleman} for a comprehensive exposition on the topic. Our principal interest in this work is to examine the interaction between the nematic defects and isotropic-to-nematic interfaces. 

The interfaces form in the process of the first order isotropic-to-nematic phase transition that can be induced either by lowering temperature (thermotropic nematics) or increasing the concentration of asymmetric molecules in a solvent (lyotropic nematics). Typically, nuclei of the nematic phase form within the isotropic phase upon lowering the temperature. The nematic nuclei are separated from the isotropic regions by phase boundaries or interfaces. The interfaces subsequently propagate converting the isotropic phase to the nematic phase in the process. In addition to motion of interfaces, the resulting dynamics of the system involves formation, annihilation and propagation of various types of defects. 

In a recent work \cite{Kim_2013}, the authors examine the interplay between the interfaces and defects that are present during phase transitions in lyotropic chromonic liquid crystals (LCLC). The principal feature of the isotropic-to-nematic phase transition in LCLC is that the interface provides an ``easy direction" for nematic anchoring on the interface which in this case forces the director field to be tangent to this phase boundary. When combined with the anchoring (boundary) conditions on the walls of the container, the topology of the nematic configuration leads to formation of structural defects. 

Our goal in the present paper is to demonstrate that the zoo of singularities observed in \cite{Kim_2013} can be described within the framework of the Landau-de Gennes model for $Q$-tensors related to the projection matrix descriptor of the nematic phase alluded to above. Critical to our modeling will be an assumption of large disparity between the values of the elastic constants appearing in the energy. In Section \ref{sec:exp} we briefly describe the experimental observations that expand on some of the results presented in \cite{Kim_2013} by incorporating scenarios of the phase transition in which the isotropic phase regions, often called ``negative tactoids" \cite{Nastishin}, \cite{Kim_2013}, or simply ``tactoids", shrink into a uniform nematic state or a state with a topological defect, depending on the winding number of the tactoid. Note here that positive tactoids, i.e., regions of a nematic surrounded by an isotropic melt, have been previously theoretically treated in \cite{Kaznacheev1}-\nocite{Kaznacheev2,Prinsen1,Prinsen2,Prinsen3}\cite{Acharya}.

In Section \ref{sec:LdG} we review the basics of the Landau-de Gennes theory and then develop our model in Section \ref{sec:main}. In Section \ref{sec:num} we describe our numerical results and compare them with experimental observations. 

\section{Experimental Results}
\label{sec:exp}
The sample configuration consisted of the ITO glass that was spin-coated with a polyimide layer, SE7511, in which the directors of disodium chromoglycate (DSCG) are aligned parallel to substrates. Subsequently, two substrates were assembled into a cell with the thickness of 2 $\mu$m. $16$ wt\% of DSCG solution was injected into the assembled cell.

\begin{figure}[htp]
        \centering
        \includegraphics[width=.3\linewidth,height=.3\linewidth]{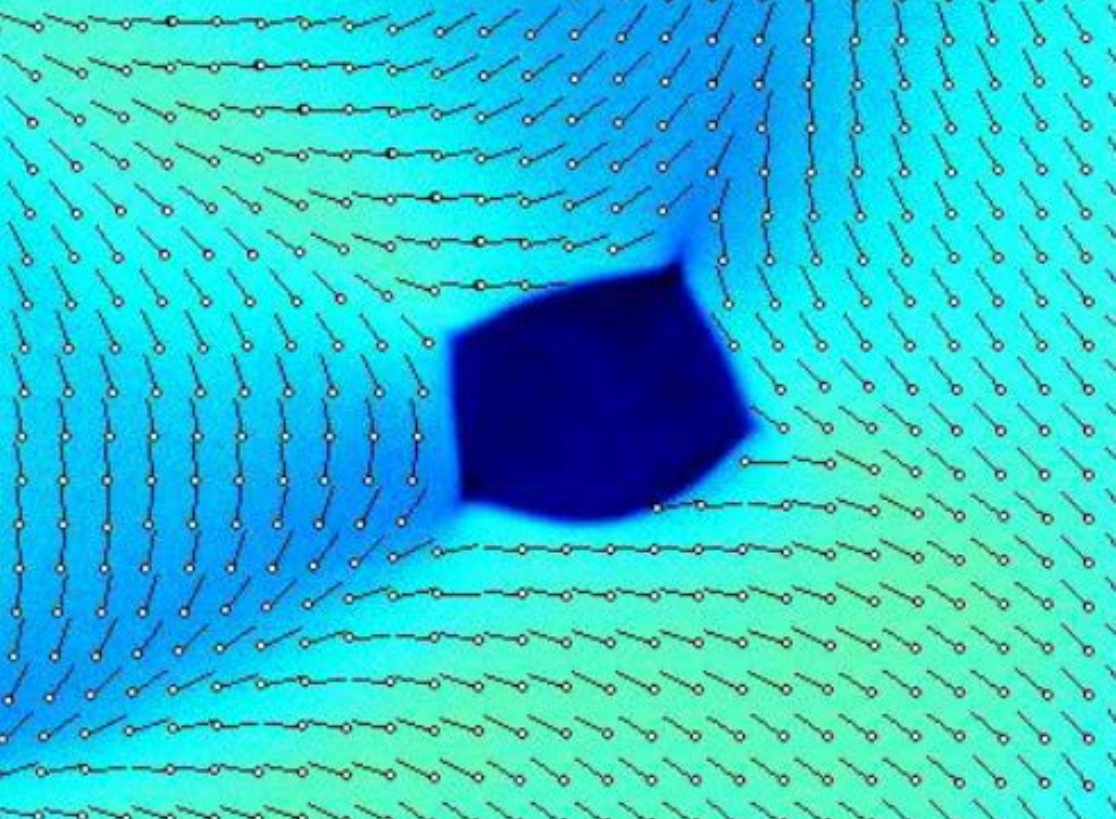}\quad
        \includegraphics[width=.3\linewidth,height=.3\linewidth]{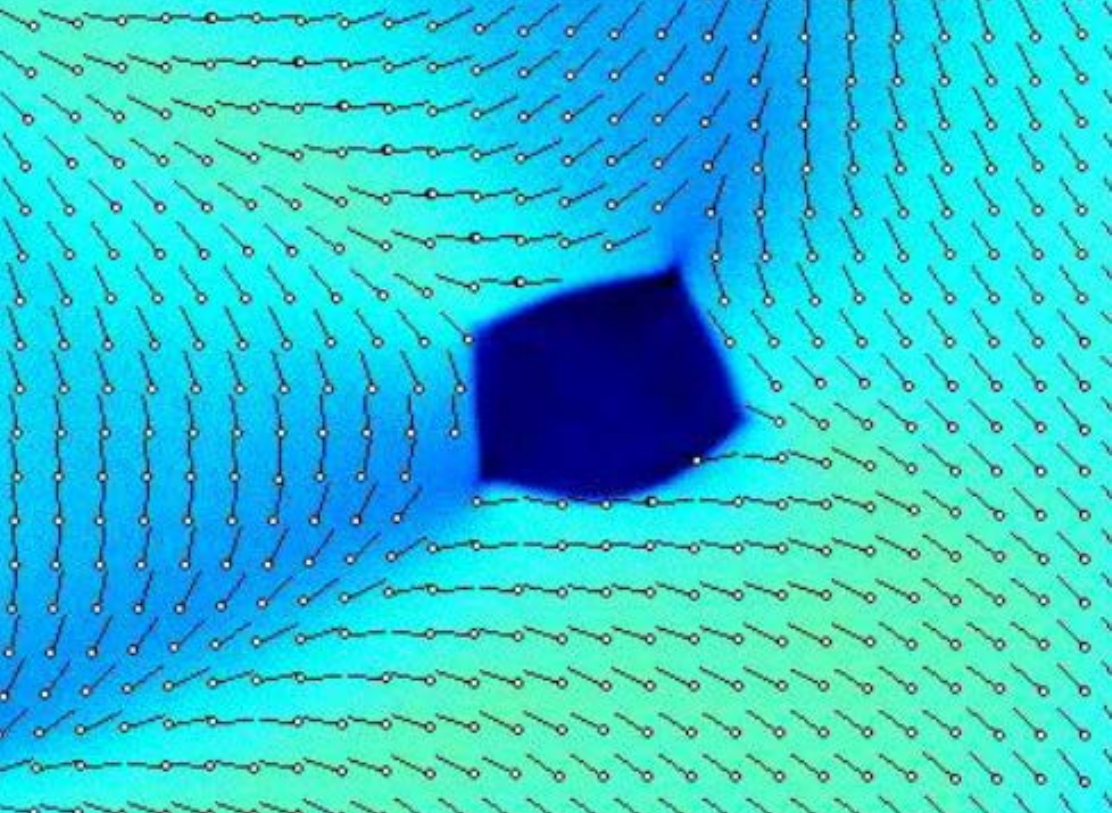}\quad \includegraphics[width=.3\linewidth,height=.3\linewidth]{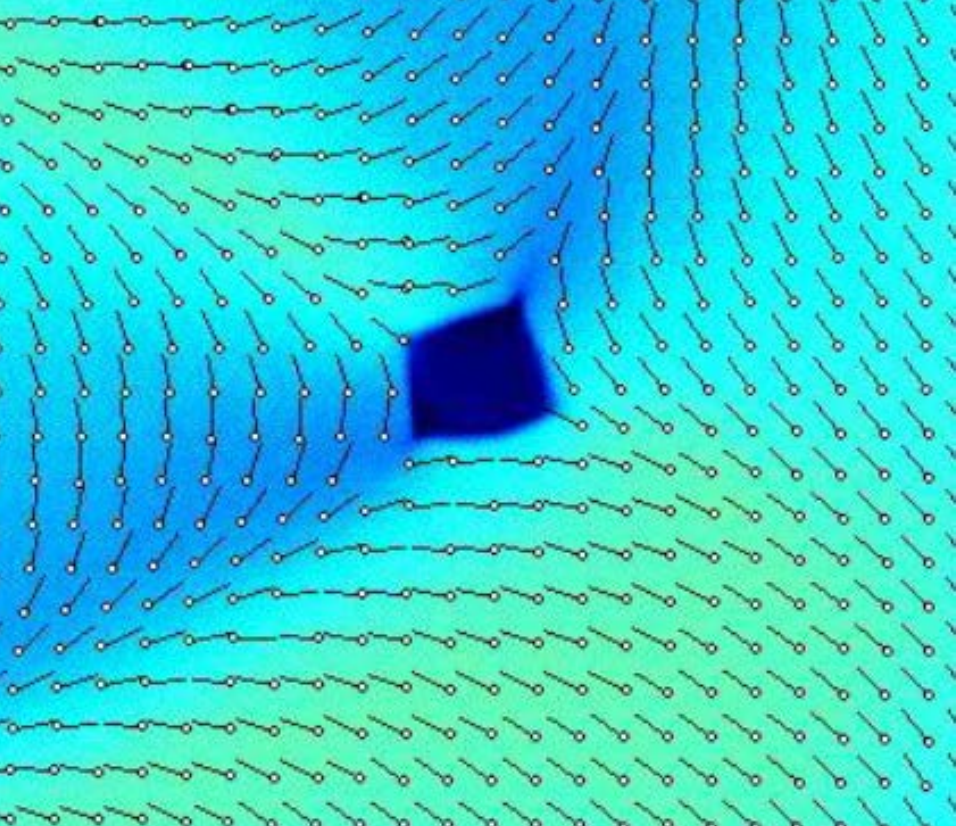}
        
        \vspace{3mm}
        
        \hspace{.3em}\includegraphics[width=.3\linewidth,height=.3\linewidth]{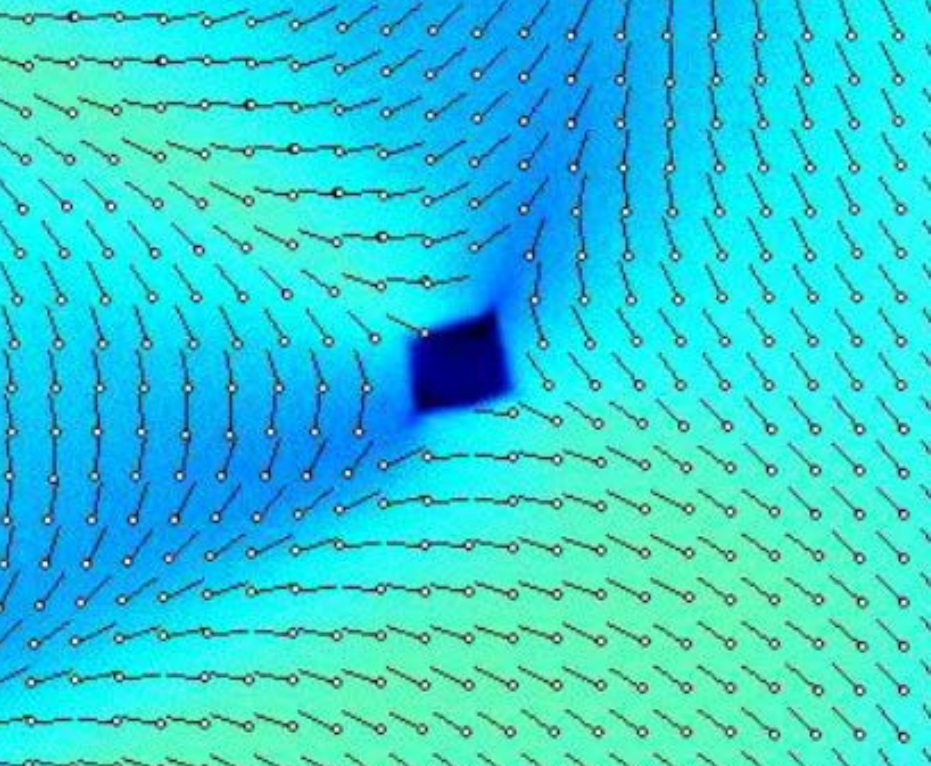}\hspace{-.3em} \quad \includegraphics[width=.3\linewidth,height=.3\linewidth]{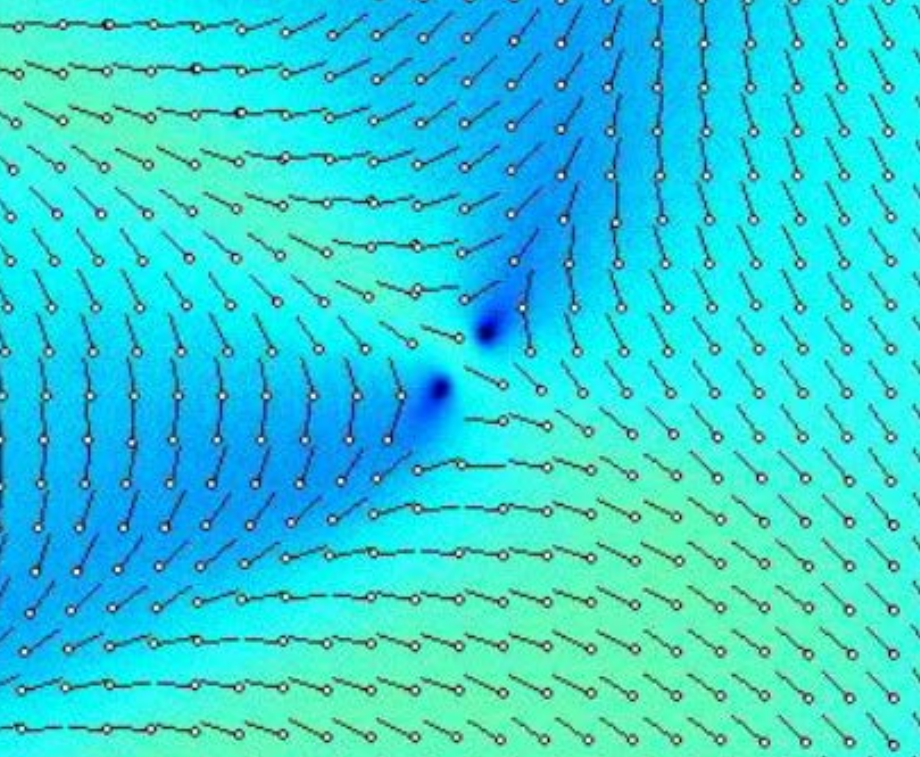}\quad \includegraphics[width=.3\linewidth,height=.3\linewidth]{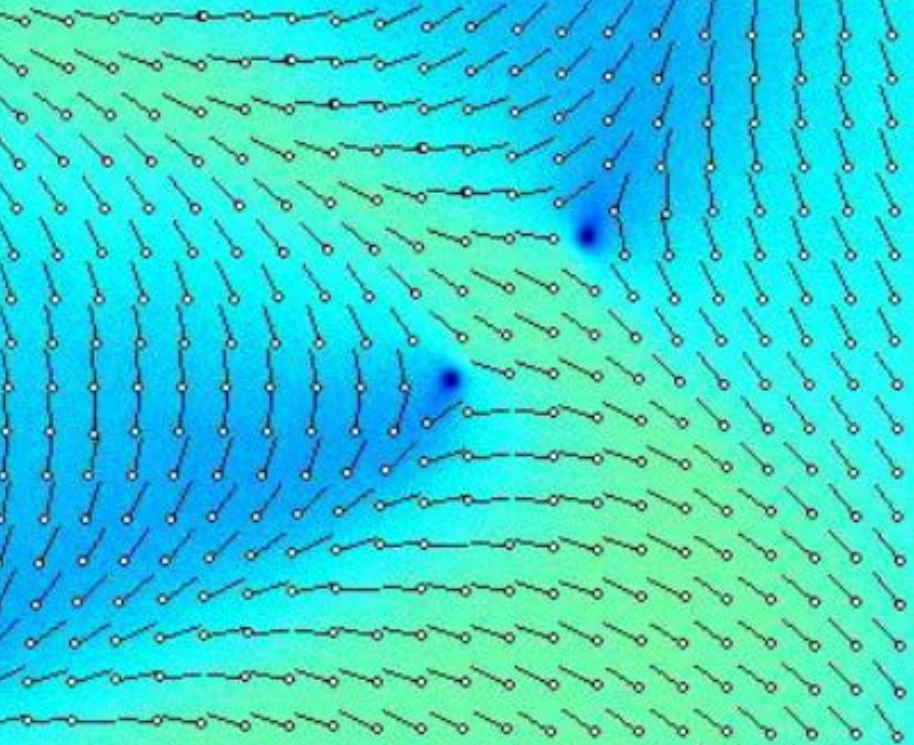}
        \caption{Experimentally observed evolution of an isotropic tactoid. The director field has the winding number $-1$ on the boundary of the tactoid. Once the tactoid disappears, it generates a vortex of degree $-1$ that subsequently splits into two degree $-1/2$ vortices.}
        \label{fig:my_label}
    \end{figure}

\begin{figure}[htp]
        \centering
        \includegraphics[width=.3\linewidth,height=.3\linewidth]{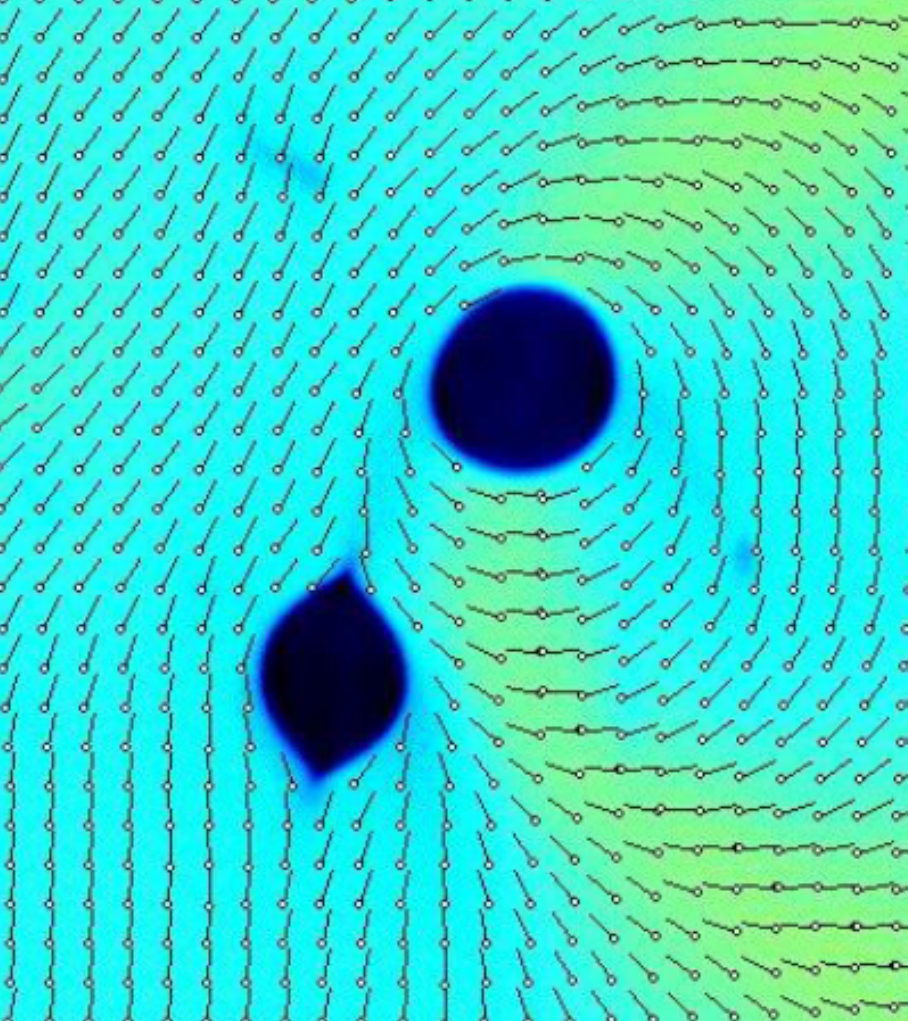}\quad
        \includegraphics[width=.3\linewidth,height=.3\linewidth]{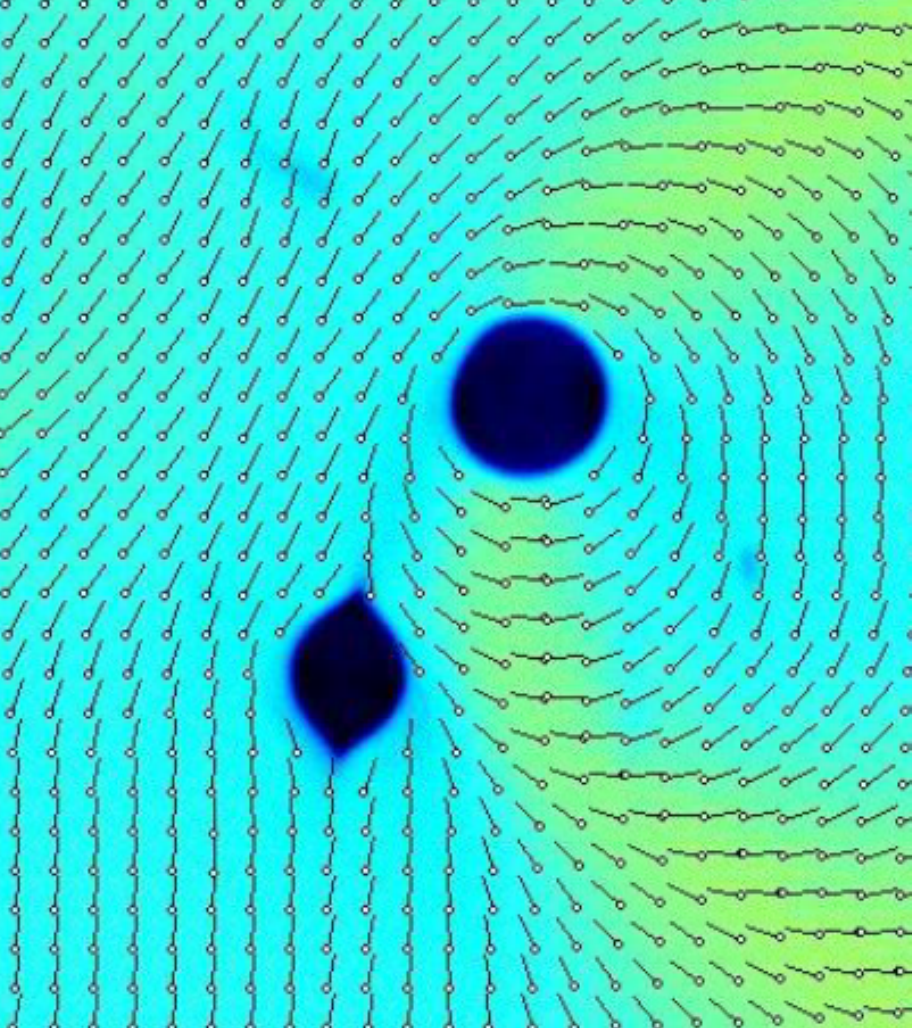}\quad \includegraphics[width=.3\linewidth,height=.3\linewidth]{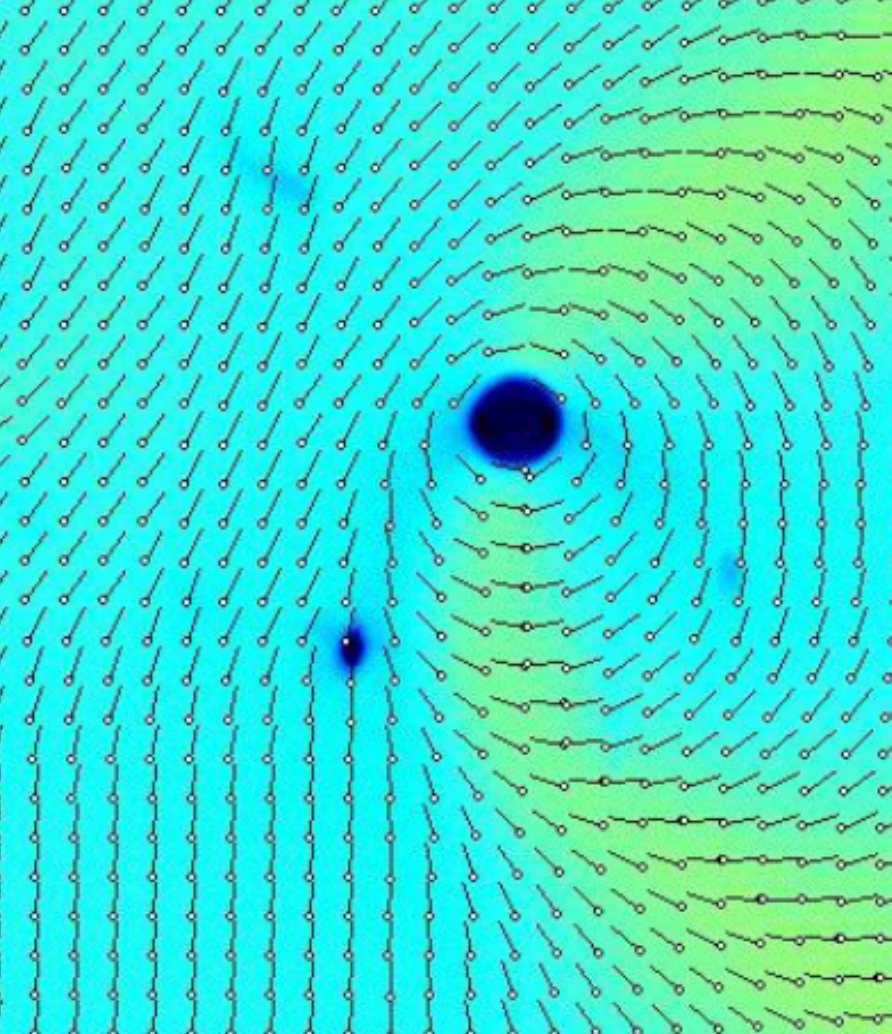}
        
        \vspace{3mm}
        
        \hspace{.3em}\includegraphics[width=.3\linewidth,height=.3\linewidth]{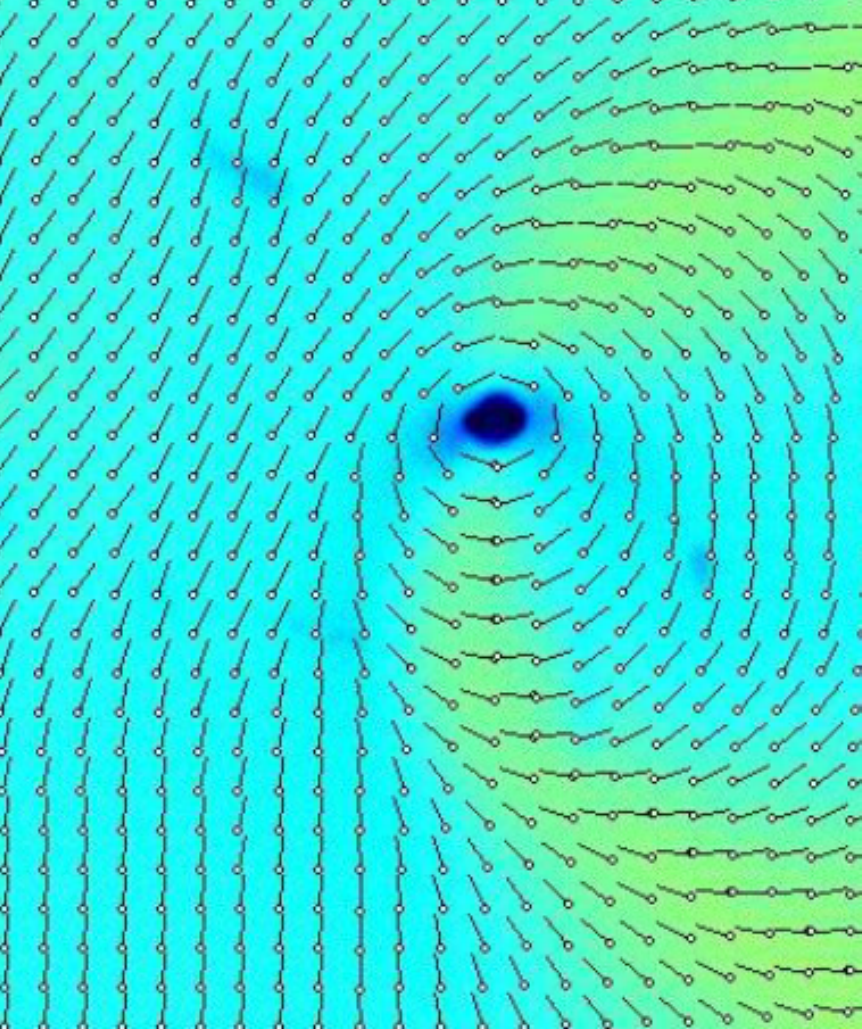}\hspace{-.3em} \quad \includegraphics[width=.3\linewidth,height=.3\linewidth]{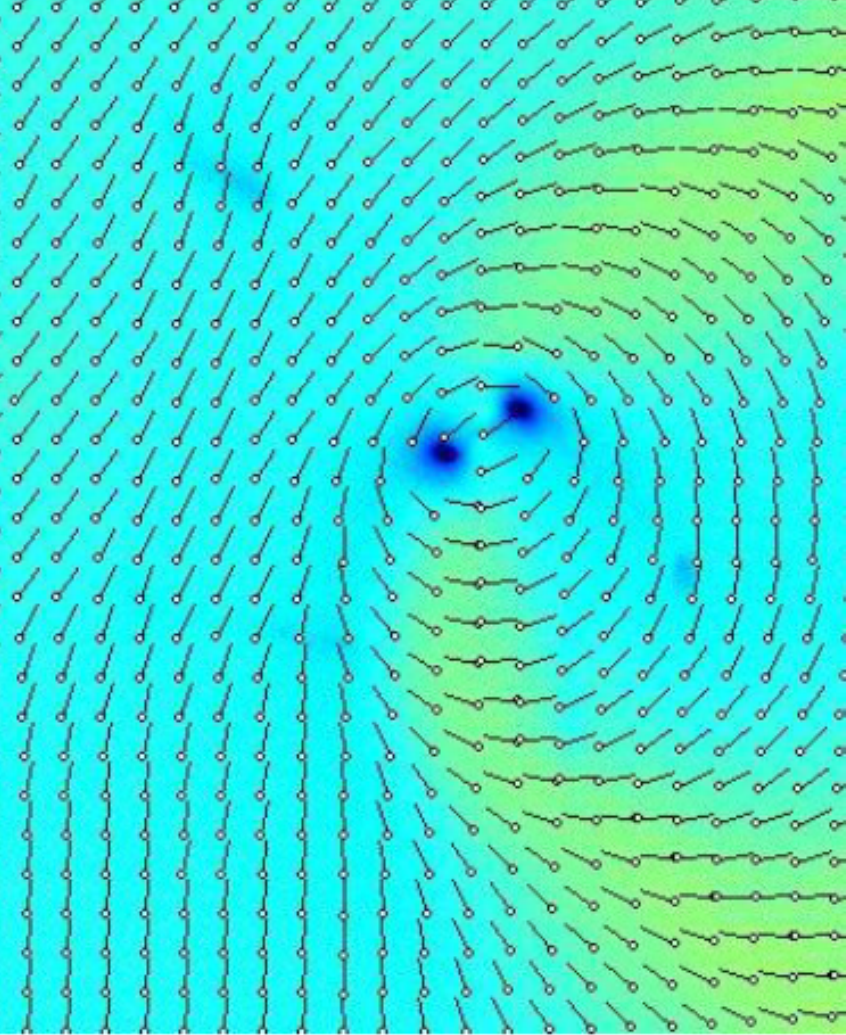}\quad \includegraphics[width=.3\linewidth,height=.3\linewidth]{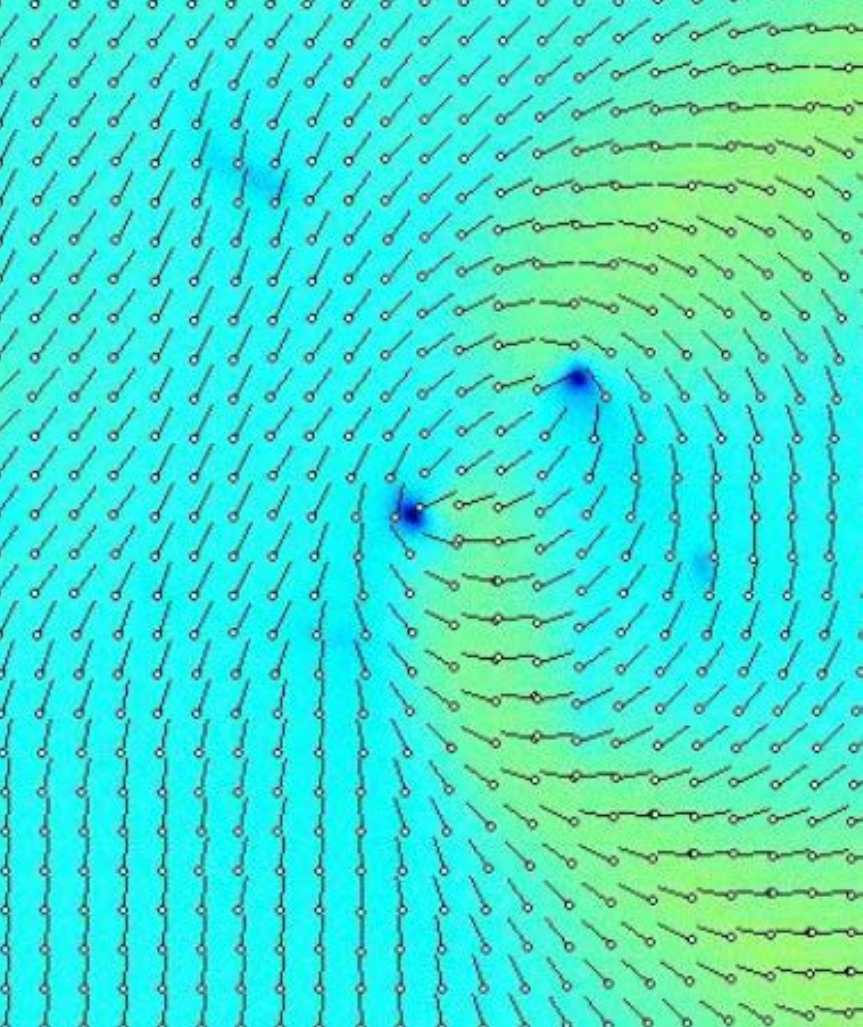}
        \caption{Experimentally observed evolution of isotropic tactoids. On each inset, the director field on the boundary of the right tactoid has the winding number $0$ while it is equal to $1$ on the boundary of the right tactoid. Once the right tactoid disappears, it generates a vortex of degree $1$ that subsequently splits into two degree $1/2$ vortices. The disappearance of the left tactoid does not lead to the formation of topological singularities.}
        \label{fig:your_label}
    \end{figure}

The cell was cooled at the rate of $-0.4^\circ C/min$. As the temperature decreased from the isotropic phase, nematic domains appeared, grew, and coalesced. When many large nematic domains coalesced, they occasionally trapped isotropic islands, or tactoids, around which the director may have either zero or nonzero winding number. The snapshots of configurations with islands having different winding number of the director on their boundary are shown in Figs. \ref{fig:my_label}-\ref{fig:your_label}. These are Polscope textures with color representing optical retardance, bars giving the orientation of the director, and circles indicating the point where orientation was measured. 
    
\section{Landau-de Gennes Model}
\label{sec:LdG}
\subsection{The $Q$-tensor}
\label{s:qtens}
Given a point $x\in\mathbb R^3$, the second moment of the orientational distribution of the rod-like nematic liquid crystal molecules near $x$ can be described by a $2$-tensor $Q(x)$ that takes the form of a $3\times  3$ symmetric, traceless matrix. By virtue of being symmetric and traceless, $Q$ has three real eigenvalues $\lambda_1,\ \lambda_2,$ and $\lambda_3$ that satisfy
\[\lambda_1+\lambda_2+\lambda_3=0\] 
and a mutually orthonormal eigenframe $\left\{\mathbf{l},\mathbf{m},\mathbf{n}\right\}$. While a more detailed overview of the theory can be found in \cite{Mottram_Newton}, below we outline the key elements that will be needed in the developments below.

Suppose that $\lambda_1=\lambda_2=-\lambda_3/2.$ Then the liquid crystal is in a {\em uniaxial nematic} state and \begin{equation}Q=-\frac{\lambda_3}{2}\mathbf{l}\otimes\mathbf{l}-\frac{\lambda_3}{2}\mathbf{m}\otimes\mathbf{m}+
\lambda_3\mathbf{n}\otimes\mathbf{n}=S\left(\mathbf{n}\otimes\mathbf{n}-\frac{1}{3}\mathbf{I}\right),\label{uniaxial}
 \end{equation}
 where $S:=\frac{3\lambda_3}{2}$ is the uniaxial nematic order parameter and $\mathbf{n}\in\mathbb{S}^2$ is the nematic director. If there are no repeated eigenvalues, the liquid crystal is said to be in a {\em biaxial nematic} state, while it is in the isotropic state when all three eigenvalues and hence $Q$ itself vanish. The isotropic state of a liquid crystal is associated, for instance, with a high temperature regime.

From the modeling perspective it turns out that the eigenvalues of $Q$ must satisfy the constraints \cite{ball2010nematic,sonnet2012dissipative}:
\begin{equation}
\label{eq:bnds}
\lambda_i\in[-1/3,2/3],\ \mathrm{for}\ i=1,2,3.
\end{equation}

\subsection{Landau-de Gennes energy}
Suppose that the bulk elastic energy density of a nematic liquid crystal is given by a frame-indifferent expression $f_e(Q,\nabla Q)$. A common expression that includes terms up to the second order in $Q$ reads
\begin{multline}
\label{elastic}
f_e(\nabla Q):=\frac{L_1}{2}{|\nabla Q|}^2+\frac{L_2}{2}Q_{ij,j}Q_{ik,k}+\frac{L_3}{2}Q_{ik,j}Q_{ij,k} \\
=\sum_{j=1}^3\left\{\frac{L_1}{2}{|\nabla Q_j|}^2+\frac{L_2}{2}\left(\dvr{Q_j}\right)^2+\frac{L_3}{2}\nabla Q_j\cdot \nabla Q_j^T\right\}.
\end{multline}
The bulk Landau-de Gennes energy density is
\begin{equation}
\label{LdG}
W(Q):=3a\,\mathrm{tr}\left(Q^2\right)-2b\,\mathrm{tr}\left(Q^3\right)+\frac{1}{4}\left(\mathrm{tr}\left(Q^2\right)\right)^2,
\end{equation}
cf. \cite{Mottram_Newton}. Here $Q_j,\,j=1,2,3$ is the $j$-th column of the matrix $Q$ and $A\cdot B=\tr{\left(B^TA\right)}$ is the dot product of two matrices $A,B\in M^{3\times3}.$ Further, the coefficient $a$ is temperature-dependent and in particular is negative for sufficiently low temperatures.
One readily checks that the form \eqref{LdG} of this potential implies that in fact $W$ depends only on the eigenvalues of $Q$, and due to the trace-free condition, therefore
 depends only on two eigenvalues. The form of $W$ guarantees that the isotropic state $Q\equiv 0$ yields a global minimum at high temperatures while a uniaxial state of the form \eqref{uniaxial} gives the minimum when temperature (i.e. the parameter $a$) is reduced below a certain critical value, cf. \cite{apala_zarnescu_01,Mottram_Newton}. We remark for future use that $W$ is bounded from below and can be made non-negative by adding an appropriate constant.

We now turn to the behavior of the nematic on the boundary of the sample. Here two alternatives are possible. First, the Dirichlet boundary conditions on $Q$ are referred to as strong anchoring conditions in the physics literature: they impose specific preferred orientations on nematic molecules on surfaces bounding the liquid crystal. An alternative is to specify the anchoring energy on the boundary of the sample; then orientations of the molecules on the boundary are determined as a part of the minimization procedure. This approach is known as {\em weak anchoring}. 

Putting the energy densities together, cf. \eqref{elastic}, \eqref{LdG}, we arrive at a Landau-de Gennes type model to be analyzed in this study, given by
\begin{equation}
\label{energy}
E[Q]:=\int_{\Omega}\left\{f_e(Q,\nabla Q)+W(Q)\right\}\,dV.
\end{equation}

\section{Model Development}
\label{sec:main}

In this section we derive a version of the Landau-de Gennes model that is appropriate for the modeling of nematic systems with disparate elastic constants. In particular, we are interested in the case when the elastic constant corresponding to splay deformations is larger than those for bend and twist so that the splay of the director is relatively expensive. Note that this situation can be found in experimental systems, such as chromonic lyotropic liquid crystals \cite{Zhou} shown in Figures \ref{fig:my_label}-\ref{fig:your_label} and in thermotropic nematics of certain molecular shape, such as dimers \cite{Greta}. To this end, let \[A:=\left\{Q\in M^{3\times3}:Q^T=Q,\,\tr{Q}=0\right\}\] and consider the Landau-de Gennes potential $W(Q)$ defined in \cref{LdG}. As long as $Q\in A$, one finds that this potential depends only on the eigenvalues of $Q$, say $\lambda_1$ and $\lambda_2$ with $\lambda_3=-(\lambda_1+\lambda_2)$, and with a slight abuse of notation we arrive at 
\[W(\lambda_1,\lambda_2)=6a\left(\lambda_1^2+\lambda_2^2+\lambda_1\lambda_2\right)+6b\lambda_1\lambda_2\left(\lambda_1+\lambda_2\right)+\left(\lambda_1^2+\lambda_2^2+\lambda_1\lambda_2\right)^2.\]

We are interested in describing the extrema of $W(\lambda_1,\lambda_2)$. Taking the derivatives of $W$ gives
\begin{align*}
    \frac{\partial W}{\partial \lambda_1}&=2\left(2\lambda_1+\lambda_2\right)\left(3a+3b\lambda_2+\left(\lambda_1^2+\lambda_2^2+\lambda_1\lambda_2\right)\right),\\
     \frac{\partial W}{\partial \lambda_2}&=2\left(\lambda_1+2\lambda_2\right)\left(3a+3b\lambda_1+\left(\lambda_1^2+\lambda_2^2+\lambda_1\lambda_2\right)\right),\\
         \frac{\partial^2 W}{\partial \lambda_1^2}&=12a+12b\lambda_2+2\left(2\lambda_1+\lambda_2\right)^2+4\left(\lambda_1^2+\lambda_2^2+\lambda_1\lambda_2\right),\\
         \frac{\partial^2 W}{\partial \lambda_2^2}&=12a+12b\lambda_1+2\left(\lambda_1+2\lambda_2\right)^2+4\left(\lambda_1^2+\lambda_2^2+\lambda_1\lambda_2\right),\\
         \frac{\partial^2 W}{\partial \lambda_1\partial\lambda_2}&=6a+12b\left(\lambda_1+\lambda_2\right)+2\left(2\lambda_1+\lambda_2\right)\left(\lambda_1+2\lambda_2\right)+2\left(\lambda_1^2+\lambda_2^2+\lambda_1\lambda_2\right).
\end{align*}
It is easy to check that the critical points of $W$ are $(\lambda_1,\lambda_2)=(0,0)$ as well as $(\lambda_1,\lambda_2)=(\lambda,-2\lambda)$, $(\lambda_1,\lambda_2)=(-2\lambda,\lambda)$, and $(\lambda_1,\lambda_2)=(\lambda,\lambda)$, where $\lambda$ solves
\begin{equation}
\label{eq:1}
    a+b\lambda+\lambda^2=0.
\end{equation}
The point $(\lambda_1,\lambda_2)=(0,0)$ corresponds to a local minimum of $W$ if $a>0$ and to a local maximum of $W$ if $a<0$. The solutions of \cref{eq:1} are given by
\[\lambda=\frac{-b\pm\sqrt{b^2-4a}}{2},\]
hence there are no nematic critical points whenever $a>b^2/4$ and $W$ has a single extremum corresponding to the global minimum at $(\lambda_1,\lambda_2)=(0,0)$. By considering second derivatives of $W$, we also find that 
\begin{equation}
    \label{eq:2}
    \lambda_m=\frac{-b-\sqrt{b^2-4a}}{2}
\end{equation}
is a point of a local minimum and
\[\lambda_s=\frac{-b+\sqrt{b^2-4a}}{2}\]
is a saddle point of $W$ if $a<b^2/4$. We conclude that the nematic minima at $(\lambda_m,-2\lambda_m)$, $(-2\lambda_m,\lambda_m)$, and $(\lambda_m,\lambda_m)$ coexist with the isotropic minimum at $(0,0)$ as long as $0<a<b^2/4$. An easy computation shows that all of these minima have the same depth of $0$ when
\begin{equation}
    \label{eq:a0}
    a=a_0:=2b^2/9.
\end{equation}
Because here we will be interested in the regime $0\leq a\leq a_0$ when the energy value in the isotropic state is greater than or equal to the minimum energy in a nematic state, we subtract 
\begin{equation}
    \label{eq:3}
    W_m:=W(\lambda_m,\lambda_m)=9ab^2-9a^2-\frac{3}{2}b^4-\frac{3}{2}b{\left(b^2-4a\right)}^\frac{3}{2}
\end{equation}
from $W$ to ensure that the global minimum value of the Landau-de Gennes energy is $0$. Hence, from now on
\[W(Q)\to W(Q)-W_m.\]
Note that decreasing $a$ from $a_0$ to $0$ corresponds to increased undercooling and a larger thermodynamic force driving the isotropic-to-nematic phase transition. 

Let $\Omega\subset\mathbb R^3$ denote the region occupied by the liquid crystal sample. To derive an expression for the elastic energy, recall first that the Oseen-Frank energy defined over director fields $n\in H^1\left(\Omega,\mathbb S^2\right)$ is given by 
\[E_{OF}[n]=\frac{1}{2}\int_\Omega\left\{K_1{\left(\dv{n}\right)}^2+K_2{\left(n\cdot\cl{n}\right)}^2+K_3{\left|n\times\cl{n} \right|}^2\right\}\,dx\]
where for simplicity we assume that the admissible competitors are subject to the same Dirichlet boundary data on $\partial\Omega$. Assuming in this paper that $K_2=K_3<K_1$, the energy can be written (up to null Lagrangian terms) as
\begin{equation}
    \label{eq:4}
    E_{OF}[n]=\frac{1}{2}\int_\Omega\left\{\tilde K_1{\left(\dv{n}\right)}^2+\tilde K_2{\left|\nabla n\right|}^2\right\}\,dx,
\end{equation}
where $\tilde K_1=K_1-K_2$ and $\tilde K_2=K_2$ are nonnegative elastic constants. We want to derive a well-posed variational problem for $Q$-valued fields such that the elastic energy for $Q$ formally reduces to \cref{eq:4} whenever $Q$ is in a nematic state that minimizes the potential energy $W$. To this end, the minimum of $W$ is achieved whenever $Q$ has eigenvalues $\lambda_m$, $\lambda_m$, and $-2\lambda_m$ so that $Q$ can be written as
\[Q=\lambda_m\left(\mathrm{I}-3\,n\otimes n\right),\]
where $n$ is a unit eigenvector corresponding to the eigenvalue $-2\lambda_m$. From this equation, 
\[n\otimes n=\frac{1}{3}\left(\mathrm{I}-Q/\lambda_m\right)\]
and it can be easily checked that 
\[{|\nabla n|}^2=\frac{1}{18}{|\nabla (Q/\lambda_m)|}^2\]
while
\[-\frac{1}{3}\dv{(Q/\lambda_m)}=\dv{\left(n\otimes n\right)}=(\dv{n})n+\nabla n\,n.\]
Making use of the fact that
\[\left(n\otimes n\right)\nabla n\,n=\left(n\cdot\nabla n\,n\right)n=\left(\nabla n^Tn\cdot n\right)n=\frac{1}{2}\left(\nabla \left(|n|^2\right)\cdot n\right)n=0,\]
it follows that 
\[(\dv{n})n=\left(n\otimes n\right)\left((\dv{n})n+\nabla n\,n\right)=-\frac{1}{9}\left(\mathrm{I}-Q/\lambda_m\right)\dv{(Q/\lambda_m)}\]
and
\[\tilde K_1{\left(\dv{n}\right)}^2+\tilde K_2{\left|\nabla n\right|}^2\sim\frac{L_1}{2}{|\nabla (Q/\lambda_m)|}^2+\frac{L_2}{2}{\left|\left(\mathrm{I}-Q/\lambda_m\right)\dv{(Q/\lambda_m)}\right|}^2\]
where $L_1:=\frac{\tilde K_2}{9}$ and $L_2:=\frac{2\tilde K_1}{81}$. Note that here we switched the indices of the elastic constants in order to conform with the standard notation (cf. \cref{elastic}).

The total energy of a nematic configuration as described within the Landau-de Gennes $Q$-tensor theory in this work will be given by
\begin{equation}
    \label{eq:5}
    F[Q]:=\int_\Omega\left(\frac{L_1}{2}{|\nabla (Q/\lambda_m)|}^2+\frac{L_2}{2}{\left|\left(\mathrm{I}-Q/\lambda_m\right)\dv{(Q/\lambda_m)}\right|}^2+w_0W(Q)\right)\,dx,
\end{equation}
where $w_0$ is a constant that has units of energy per unit volume.
\begin{remark}
Expanding the second elastic term in \cref{eq:5} we obtain 
\[{\left|\left(\mathrm{I}-Q/\lambda_m\right)\dv{(Q/\lambda_m)}\right|}^2=\frac{1}{\lambda_m^2}{\left|\dv{Q}\right|}^2-\frac{2}{\lambda_m^3}Q\,\dv{Q}\cdot\dv{Q}+\frac{1}{\lambda_m^4}{\left|Q\,\dv{Q}\right|}^2.\]
Comparing the terms appearing in this expression with the elastic invariants of the generalized Landau-de Gennes theory in \cite{Longa}, we see that these terms correspond to $L_2^{(2)}$-, $L_3^{(3)}$-, and $L_6^{(4)}$-invariants.
\end{remark}
Now let $l>0$ denote a characteristic length of the problem and set
\[\varepsilon:=\frac{L_1}{L_2},\quad \gamma:=\frac{4L_2}{w_0l^2\lambda_m^4}.\]
Before proceeding further, we introduce the following rescalings
\begin{equation}
    \label{eq:nd}
    \bar x=\frac{x}{l},\quad\bar Q=-\frac{Q}{\lambda_m},\quad\bar F=\frac{Fl}{L_2},\quad \bar a=\frac{a}{\lambda_m^2},\quad \bar b=-\frac{b}{\lambda_m},\quad \bar W_m=\frac{W_m}{\lambda_m^4}
\end{equation}
and drop the bar notation to obtain
\begin{equation}
    \label{eq:6}
    F[Q]=\int_\Omega\left(\frac{\varepsilon}{2}{|\nabla Q|}^2+\frac{1}{2}{\left|\left(\mathrm{I}+Q\right)\dv{Q}\right|}^2+\frac{1}{4\gamma}W(Q)\right)\,dx,
\end{equation}
where
\begin{equation}
    \label{eq:7}
    W(Q)=3a\left(\tr{Q^2}\right)-2b\left(\tr{Q^3}\right)+\frac{1}{4}\left(\tr{Q^2}\right)^2-W_m.
\end{equation}
In this scaling, the potential $W$ given by \eqref{eq:7} is now minimized by any symmetric traceless matrix with eigenvalues $-1,\ -1,\ 2$, and the global minimum value of $W$ is equal to zero. 

In our simulations we consider a simplified form of a $Q$-tensor that can be obtained via a dimension reduction procedure for thin nematic films \cite{GoMoSt}. In the corresponding ansatz, one eigenvector of admissible $Q$-tensors must be perpendicular to the plane of the film. Then in the system of coordinates in which the normal to the film is parallel to the $z$-axis, the $Q$-tensor is independent of $z$ and can be written \cite{BPP} as
\begin{equation}
    \label{bpp}
    Q(x,y)=\left(\begin{array}{ccc}
      \beta(x,y)+u_1(x,y) & u_2(x,y) & 0 \\
      u_2(x,y) & \beta(x,y)-u_1(x,y) & 0 \\
      0 & 0 & -2\beta(x,y)
    \end{array}
    \right),
\end{equation}
for a scalar-valued function $\beta$ and vector-valued function $u=(u_1,u_2)$.
Let
\[    U=\left(\begin{array}{cc}
      u_1 & u_2 \\
      u_2 & -u_1
    \end{array}
    \right),\]
so that
\[\left(\begin{array}{cc}
      \beta+u_1 & u_2 \\
      u_2 & \beta-u_1
    \end{array}
    \right)=\beta\mathrm{I}_2+U,\]
where $\mathrm{I}_2$ is the $2\times2$ identity matrix. In terms of $\beta$ and $u$ the contributions to the energy density are as follows
\[\frac{\varepsilon}{2}{|\nabla Q|}^2=\varepsilon\left({|\nabla u|}^2+3{|\nabla\beta|}^2\right),\]
\[
    \frac{1}{2}{\left|\left(\mathrm{I}+Q\right)\dv{Q}\right|}^2=\frac{1}{2}{\left|\left((\beta+1)\mathrm{I}_2+U\right)\left(\nabla\beta+\dv{U}\right)\right|}^2
    \]
and
\[\frac{1}{4\gamma}W(u,\beta)=\frac{1}{4\gamma}\left({\left({|u|}^2+3\beta^2\right)}^2-12b\beta\left({|u|}^2-\beta^2\right)+6a\left({|u|}^2+3\beta^2\right)\right).\]
Suppose that the film occupies the region $\omega\in\mathbb R^2$. The system of Euler-Lagrange PDEs corresponding to the energy functional \cref{eq:6} then takes the form
\begin{multline}\label{u1}
    -\dv{\left\{2\varepsilon\nabla u_1+\sigma_3{\left((\beta+1)\mathrm{I}_2+U\right)}^2\left(\nabla\beta+\dv{U}\right)\right\}} \\
    +\sigma_3\left((\beta+1)\mathrm{I}_2+U\right)\left(\nabla\beta+\dv{U}\right)\cdot\left(\nabla\beta+\dv{U}\right) \\
    +\frac{1}{\gamma}\left({|u|}^2+3\beta^2-6b\beta+3a\right)u_1=0,
\end{multline}
\begin{multline}\label{u2}
    -\dv{\left\{2\varepsilon\nabla u_2+\sigma_1{\left((\beta+1)\mathrm{I}_2+U\right)}^2\left(\nabla\beta+\dv{U}\right)\right\}} \\
    +\sigma_1\left((\beta+1)\mathrm{I}_2+U\right)\left(\nabla\beta+\dv{U}\right)\cdot\left(\nabla\beta+\dv{U}\right) \\
    +\frac{1}{\gamma}\left({|u|}^2+3\beta^2-6b\beta+3a\right)u_2=0,
\end{multline}
\begin{multline}\label{beta}
    -\dv{\left\{6\varepsilon\nabla \beta+{\left((\beta+1)\mathrm{I}_2+U\right)}^2\left(\nabla\beta+\dv{U}\right)\right\}} \\
    +\left((\beta+1)\mathrm{I}_2+U\right)\left(\nabla\beta+\dv{U}\right)\cdot\left(\nabla\beta+\dv{U}\right) \\
    +\frac{3}{\gamma}\left({|u|}^2(\beta-b)+3\beta\left(\beta^2+b\beta+a\right)\right)=0,
\end{multline}
where \[\sigma_1=\left(\begin{array}{cc}
      0 & 1 \\
      1 & 0
    \end{array}
    \right) \quad \mathrm{and} \quad \sigma_3=\left(\begin{array}{cc}
      1 & 0 \\
      0 & -1
    \end{array}
    \right)\] 
    are the Pauli matrices.
    
    When the liquid crystal is in the energy minimizing nematic state while the director $n=(\cos{\theta},\sin{\theta},0)$ lies in the plane of the film, we have
    \begin{multline*}
        \left(\begin{array}{ccc}
      \cos^2{\theta} & \cos{\theta}\sin{\theta} & 0 \\
      \cos{\theta}\sin{\theta} & \sin^2{\theta} & 0 \\
      0 & 0 & 0
    \end{array}
    \right)=n\otimes n=\frac{1}{3}(I+Q)\\=\frac{1}{3}\left(\begin{array}{ccc}
      1+\beta+u_1 & u_2 & 0 \\
      u_2 & 1+\beta-u_1 & 0 \\
      0 & 0 & 1-2\beta
    \end{array}
    \right).
    \end{multline*}
    It follows from this computation that 
    \begin{equation}
        \label{eq:wind}
        u=3\left(\cos^2{\theta}-\frac{1}{2},\cos{\theta}\sin{\theta}\right)=\frac{3}{2}\left(\cos{2\theta},\sin{2\theta}\right), \quad \beta=1/2,
    \end{equation}
    hence the vector $u$ winds twice as fast along a given curve in $\bar\omega$, compared to the director $n$ when $n$ is planar.
       
   More generally, the ansatz \cref{bpp} splits the nematic component of the minimal set of the bulk energy $W$ into two disconnected components
    \[C^N_1:=\left\{(u,\beta)\in\mathbb R^2\times\mathbb R:u=0,\beta=-1\right\}\]
    and
    \[C^N_2:=\left\{(u,\beta)\in\mathbb R^2\times\mathbb R:|u|=3/2,\beta=1/2\right\}.\]
    Here the first component corresponds to a constant nematic state with the director perpendicular to the surface of the film. The second component includes all configurations with the director lying in the plane of the film---these configurations can be nontrivial and, in particular, they include the director fields that carry a nonzero winding number. 
    
    Since we are interested in regimes when both nematic and isotropic phases coexist, we recall that whenever 
    \[0\leq a\leq\frac{2b^2}{9},\] the set 
    \[\left\{(u,\beta)\in\mathbb R^2\times\mathbb R:u=0,\beta=0\right\}\]
    yields a local minimum of $W$ corresponding to the isotropic phase. The corresponding local minimum energy value is greater than or equal to the global minimum value of $W$. When $a=\frac{2b^2}{9}$,
    the set 
    \[C^I:=\left\{(u,\beta)\in\mathbb R^2\times\mathbb R:u=0,\beta=0\right\}\]
    gives the third connected component of the minimal set of the bulk energy.
    
    Given that $W$ has a multi-component minimal set, the energy \cref{eq:6} is of Allen-Cahn-type as long as the factor in front of $W$ is large. To this end, in what follows we assume that $\gamma=\varepsilon$ and $\varepsilon>0$ is small. We will consider gradient flow dynamics associated with this model, but before we proceed further it is worth comparing our present situation to the more familiar one of gradient flow dynamics for a multi-well Allen-Cahn type potential with diffusion given simply by the Laplacian. In this scaling, when the diffusion is given by the Laplacian, formal asymptotics suggest that for Allen-Cahn dynamics one should expect an interface propagating by curvature flow that separates the different phases defined by components of the zero set of the potential. In the scalar setting of Allen-Cahn where a double well potential vanishes at two points, there are by now numerous rigorous proofs of this fact based on maximum principles, barriers and/or comparison principles, see e.g. \cite{XC,DM,ESS,Il} as well as the energetic argument in the radial setting in \cite{BK}. For the vector setting of time-dependent Allen-Cahn, formal asymptotics based on multiple time-scale expansions again suggest that mean curvature flow emerges as the governing equation for the interface, \cite{RSK1,RSK2}. 
    
    What distinguishes the dynamics in the present study, however, is the anisotropy of the diffusive terms indicated by the presence of the divergence terms in \eqref{u1}--\eqref{beta}. While we again anticipate that in the regime $\varepsilon\ll 1$ an interface separating the different phases of the $Q$ tensor will evolve by a law involving curvature, the process will be significantly affected by the interaction between the director associated with $Q$ and the normal to the interface.
    
    Our present model is closely related to the investigation in \cite{GNSV} of the $\varepsilon\to 0$ asymptotics for a director-like model based on an $\mathbb{R}^2$-valued order parameter $u$. There the elastic energy is similarly anisotropic and is coupled to a Chern-Simons-Higgs-type potential $|u|^2(|u|^2-1)^2$. Thus, the structure of the energy functional in \cite{GNSV} is similar to the setup proposed here in that it involves a potential with minima at the isotropic and a nematic state as well as the elastic terms that are quadratic in the gradient of the order parameter field. The term penalizing splay deformations in \cite{GNSV} dominates other elastic terms so that the divergence of the director is very expensive. The principal difference between the model considered in \cite{GNSV}, which from now on we will refer to as the CSH-director model,  and the present work is that here we consider non-orientable tensor fields, while the admissible fields in \cite{GNSV} are orientable. 
    
    Our numerical results indicate that similarity between the models leads to similar properties of critical points. In particular, one of the principal observations in \cite{GNSV} is that the director field is parallel to the interface in its immediate vicinity when the parameter $\varepsilon$ is small. The same behavior is exhibited by the numerically obtained critical points of the Landau-de Gennes energy \cref{eq:5}, as will be demonstrated in the next section. 
    
    We introduce dynamics into the problem by assuming that evolution of the isotropic-to-nematic transition proceeds via gradient flow
    \[\mu_uu_t=-\frac{\delta F}{\delta u}, \quad \mu_\beta\beta_t=-\frac{\delta F}{\delta \beta},\]
    where $\mu_u>0$ and $\mu_\beta>0$ are inverses of the scalar mobilities of $u$ and $\beta$, respectively. This gives the following systems of PDEs
    \begin{multline}
    \label{eq:gs1}
    \mu_u u_{1t}=\dv{\left\{2\varepsilon\nabla u_1+\sigma_3{\left((\beta+1)\mathrm{I}_2+U\right)}^2\left(\nabla\beta+\dv{U}\right)\right\}} \\
    -\sigma_3\left((\beta+1)\mathrm{I}_2+U\right)\left(\nabla\beta+\dv{U}\right)\cdot\left(\nabla\beta+\dv{U}\right) \\
    -\frac{1}{\gamma}\left({|u|}^2+3\beta^2-6b\beta+3a\right)u_1,
\end{multline}
\begin{multline}
   \label{eq:gs2}
    \mu_u u_{2t}=\dv{\left\{2\varepsilon\nabla u_2+\sigma_1{\left((\beta+1)\mathrm{I}_2+U\right)}^2\left(\nabla\beta+\dv{U}\right)\right\}} \\
    -\sigma_1\left((\beta+1)\mathrm{I}_2+U\right)\left(\nabla\beta+\dv{U}\right)\cdot\left(\nabla\beta+\dv{U}\right) \\
    -\frac{1}{\gamma}\left({|u|}^2+3\beta^2-6b\beta+3a\right)u_2,
\end{multline}
\begin{multline}
   \label{eq:gs3}
    \mu_\beta\beta_t=\dv{\left\{6\varepsilon\nabla \beta+{\left((\beta+1)\mathrm{I}_2+U\right)}^2\left(\nabla\beta+\dv{U}\right)\right\}} \\
    -\left((\beta+1)\mathrm{I}_2+U\right)\left(\nabla\beta+\dv{U}\right)\cdot\left(\nabla\beta+\dv{U}\right) \\
    -\frac{3}{\gamma}\left({|u|}^2(\beta-b)+3\beta\left(\beta^2+b\beta+a\right)\right)=0.
\end{multline}
    
    In order to simulate experimentally available configurations, we will set the initial configuration of $(u,\beta)$ to consist of an isotropic region $\omega_I$ embedded in a nematic phase corresponding to the component $C_2^N$ of the minimal set of $W$. That is, we will assume that the initial condition is an appropriate mollification of $(u,\beta)$ such that 
    \begin{equation}
        \label{eq:ic}
        (|u|,\beta)=(3/2, 1/2)\chi_{\omega\backslash\omega_I}.
    \end{equation}
    Here $\chi_{\omega\backslash\omega_I}$ is the characteristic function of the region $\omega\backslash\omega_I$ occupied by the nematic phase. We also impose boundary conditions on $\partial\omega$ with values in $C_2^N$; we will assume that the boundary data for $u$ may have a nonzero winding number.
    
    In the next section we present the result of simulations  for the system \cref{eq:gs1}-\cref{eq:gs3} and compare the numerical outcomes to experimental observations. We remind the reader that our goal in the present paper is to demonstrate numerically that behavior of interfaces and singularities observed in experiments with chromonic liquid crystals can be qualitatively recovered within the framework of the Landau-de Gennes model. We leave for future work both formal and rigorous analysis of our results, as well as any quantitative considerations and associated modifications of the model.
    
    \section{Simulations vs Experiment}
    \label{sec:num}
    The simulations in this section are performed using COMSOL {Multiphysics\textregistered} \cite{comsol} with the domain $\omega$ taken to be a disk of nondimensional radius $1/4$, centered at the origin. In what follows, we set dimensional $b=1$ and let $\gamma=\varepsilon=0.06$. The initial data in all simulations 
    is specified by choosing $\omega_I=\left\{(x,y)\in\omega:r(x,y)\leq\frac{1}{6}\right\}$ in \cref{eq:ic}. Both here and below $\left(r(x,y),\theta(x,y)\right)$ are polar coordinates of the point $(x,y)\in\mathbb R^2$.
    
    Recall that when $a$ and $b$ are related via \cref{eq:a0} (either in dimensional or nondimensional setting), the energies of the nematic and isotropic states are the same. Then, if in dimensional variables we set
    \[a=\frac{2b^2}{9}-\alpha,\]
    the parameter $\alpha\in\left(0,2b^2/9\right)$ measures the degree of {\em undercooling} in the system. That is, when $\alpha$ increases, the bulk energy density of the isotropic state grows with respect to that of the nematic state and thermodynamic forces driving the isotropic-to-nematic transition become stronger. Setting $\bar\alpha=\alpha/\lambda^2$ and dropping the bar, the nondimensional $a$ and $b$ are given by
    \begin{equation}
        \label{eq:alpha}
        a=\frac{4(2-9\alpha)}{{\left(3+\sqrt{1+36\alpha}\right)}^2},\quad b=\frac{6}{3+\sqrt{1+36\alpha}}.
    \end{equation}
    First, we verify the conjecture that the behavior of interfaces in our Landau-de Gennes system of PDEs with highly anisotropic elastic constants resembles that of interfaces in the related CSH-director system of \cite{GNSV}. To this end, we assume that $a$ is given by \cref{eq:a0} so that $\alpha=0$ and both the nematic and the isotropic states are the global minimum of the Landau-de Gennes potential $W$. We simulate the system \cref{eq:gs1}-\cref{eq:gs3} subject to the boundary conditions
    \[\beta|_{\partial\omega}=\frac{1}{2},\quad u|_{\partial\omega}=\frac{3}{2}(-\cos{2\theta},\sin{2\theta}).\]
    Taking into account \cref{eq:wind}, the director field on the boundary can be chosen as
    \[n|_{\partial\omega}=(-\cos{\theta},\sin{\theta}),\]
    where the topological degree of $n$ is equal to $-1$ on $\partial\omega$.
    
\begin{figure}
        \centering
        \includegraphics[width=.6\linewidth]{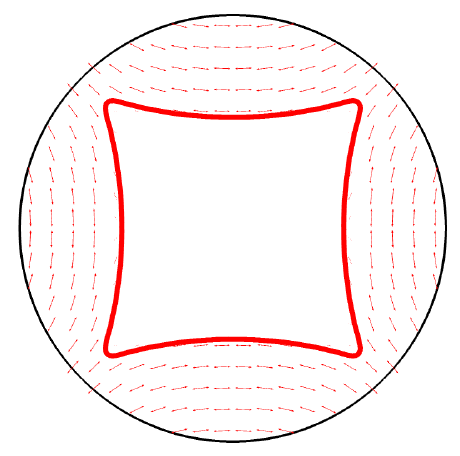}
        \caption{Simulated degree $-1$ tactoid. Here the director field $n$ is set to be equal to $(-\cos{\theta},\sin{\theta})$ on the boundary of the disk and $\theta$ is a polar angle. The thick red line indicates the position of the interface.} 
        \label{fig:paralel}
    \end{figure}

    The results of the gradient flow simulation for the degree $-1$ boundary data are shown in Fig. \ref{fig:paralel}. An initially circular interface is seen to evolve into the shape given in the figure by a thick red curve that is also a contour line of $\beta=1/4$. 
    The system in Fig. \ref{fig:paralel} has reached a steady state that is very similar to what is observed for the CSH-director model in \cite{GNSV}. Indeed, for small $\varepsilon$, the energy of the isotropic/nematic configuration in both cases should include a penalty for divergence of the director in the nematic phase, as well as the cost for the perimeter of the interface. The anisotropy of the elastic constants also forces the director to align with the interface. The resulting director configuration and the shape of the interface are coupled; in particular, unlike the standard curvature flow, shrinking the perimeter and, thus, size of the isotropic region leads to unbounded growth of the elastic energy. The competition between the perimeter and elastic energy contributions enforces the equilibrium between the two phases observed in Fig. \ref{fig:paralel}.  
    
    One is then tempted to ask whether there is any difference between the Landau-de Gennes and the CSH-director model \cite{GNSV} in terms of how the director field extracted from solutions of these two models should behave as $\varepsilon\to0$. We expect that the difference would manifest itself when the size of an isotropic region is small enough so that this region can be thought of as a core of a topological vortex. 
    
    Suppose, for example, that the interface is a single smooth closed curve. Due to the director tangency condition on the interface and because the director field is orientable, the director has the winding number $1$ around the interface. When the island shrinks to a small size and because the director is $\mathbb S^1$-valued in the nematic phase, topological constraints would keep the isotropic island from disappearing completely in order to prevent the nematic configuration from having an infinite energy. The resulting degree $1$ vortex is stable in the CSH-director setting and would persist for a finite time, perhaps until it annihilates with another vortex of the opposite sign or collides with the boundary of $\omega$.  
    
    Now recall that within the Landau-de Gennes theory, a nematic state is described by a $Q$-tensor that is a translation and dilation of a projection matrix $n\otimes n$. The field $n\otimes n$ is {\em not} orientable and can be associated with an element of a projective plane $\mathbb {RP}^2$. In other words, by working with $n\otimes n$ instead of $n$, we identify the opposite directions $-n$ and $n$ as being the same. As the result, the smallest ``quantum" of the winding number for a nematic tensor is $1/2$. Hence the degree $1$ vortex in the director description contains two ``quanta" of degree if we switch to the Landau-de Gennes framework. It is a well-established fact that a higher degree vortex is unstable with respect to splitting into several vortices of smaller degrees since the energetic cost of a degree $d$ vortex is proportional to $d^2$, \cite{BBH}. We expect that the degree $1$ CSH vortex would split into two degree $1/2$ vortices when considered in the sense of Landau-de Gennes. Our experimental observations conform to the latter picture; hence the description that disposes with orientability is necessary when considering an isotropic-to-nematic phase transition problem.
    
    We now conduct numerical experiments to test whether evolution of tactoids observed in physical experiments can be captured within the Landau-de Gennes model. 
    
    \subsection{Degree $1$ tactoid}

    Here we suppose that the director is parallel to the boundary of the disk $\omega$, e.g., 
        \[n|_{\partial\omega}=(-\sin{\theta},\cos{\theta}),\]
    so that the topological degree of the orientable director field $n$ is equal $1$ on $\partial\omega$. In the non-orientable, $Q$-tensor setting the corresponding condition can be expressed as
        \[\beta|_{\partial\omega}=\frac{1}{2},\quad u|_{\partial\omega}=-\frac{3}{2}(\cos{2\theta},\sin{2\theta}).\]    
     If the simulations are conducted at zero undercooling, when $a$ is given by \cref{eq:a0}, we observed both here and for the Ginzburg-Landau-type model \cite{GNSV} that, similar to Fig. \ref{fig:paralel}, the isotropic domain evolving via gradient flow shrinks down to a certain size and then stabilizes. In order to drive this size down, we enforce larger undercooling by choosing $\alpha=0.2.$
    
\begin{figure}[htp]
        \centering
        \includegraphics[width=.4\linewidth,height=.4\linewidth]{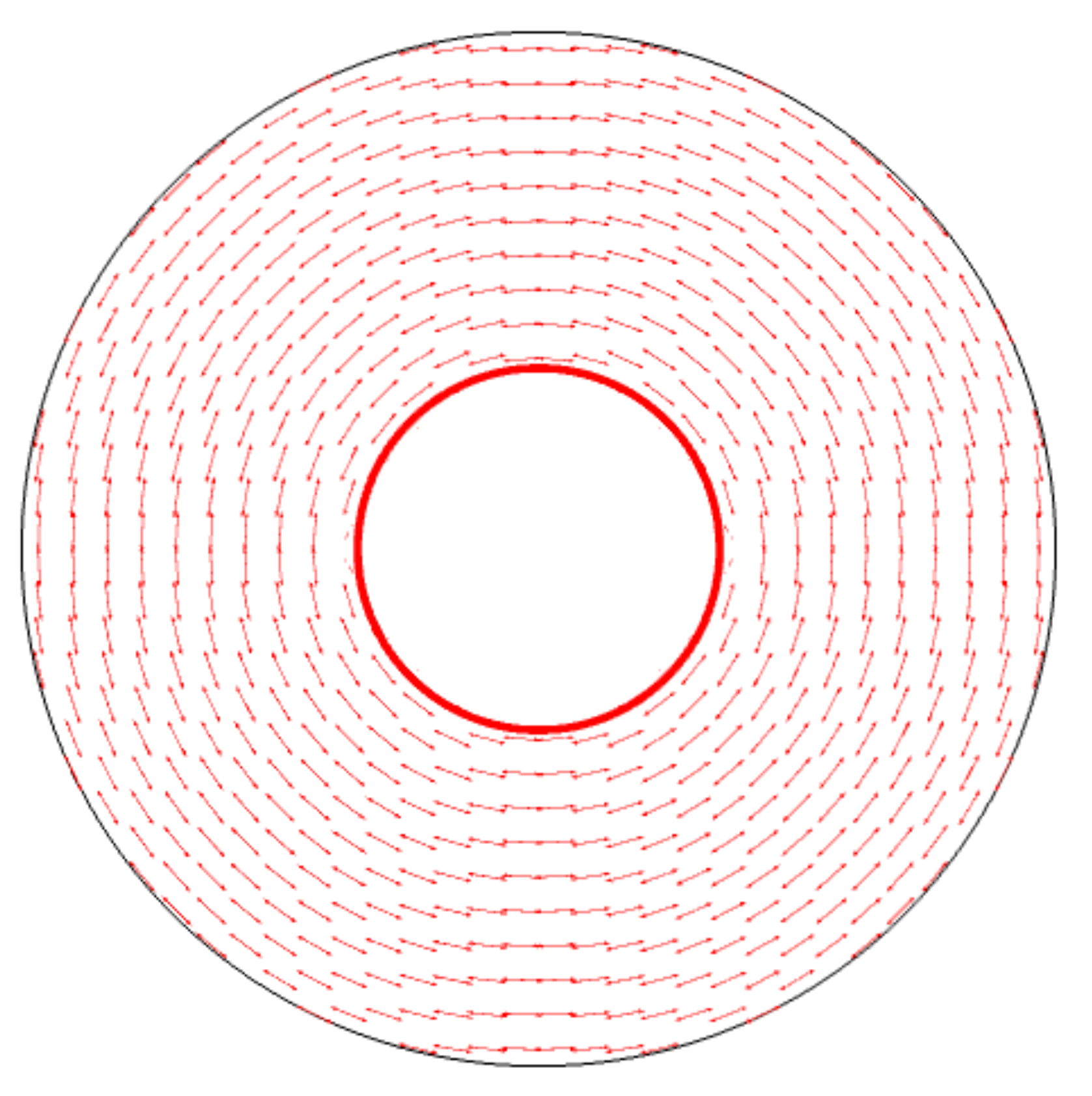}\quad
        \includegraphics[width=.4\linewidth,height=.4\linewidth]{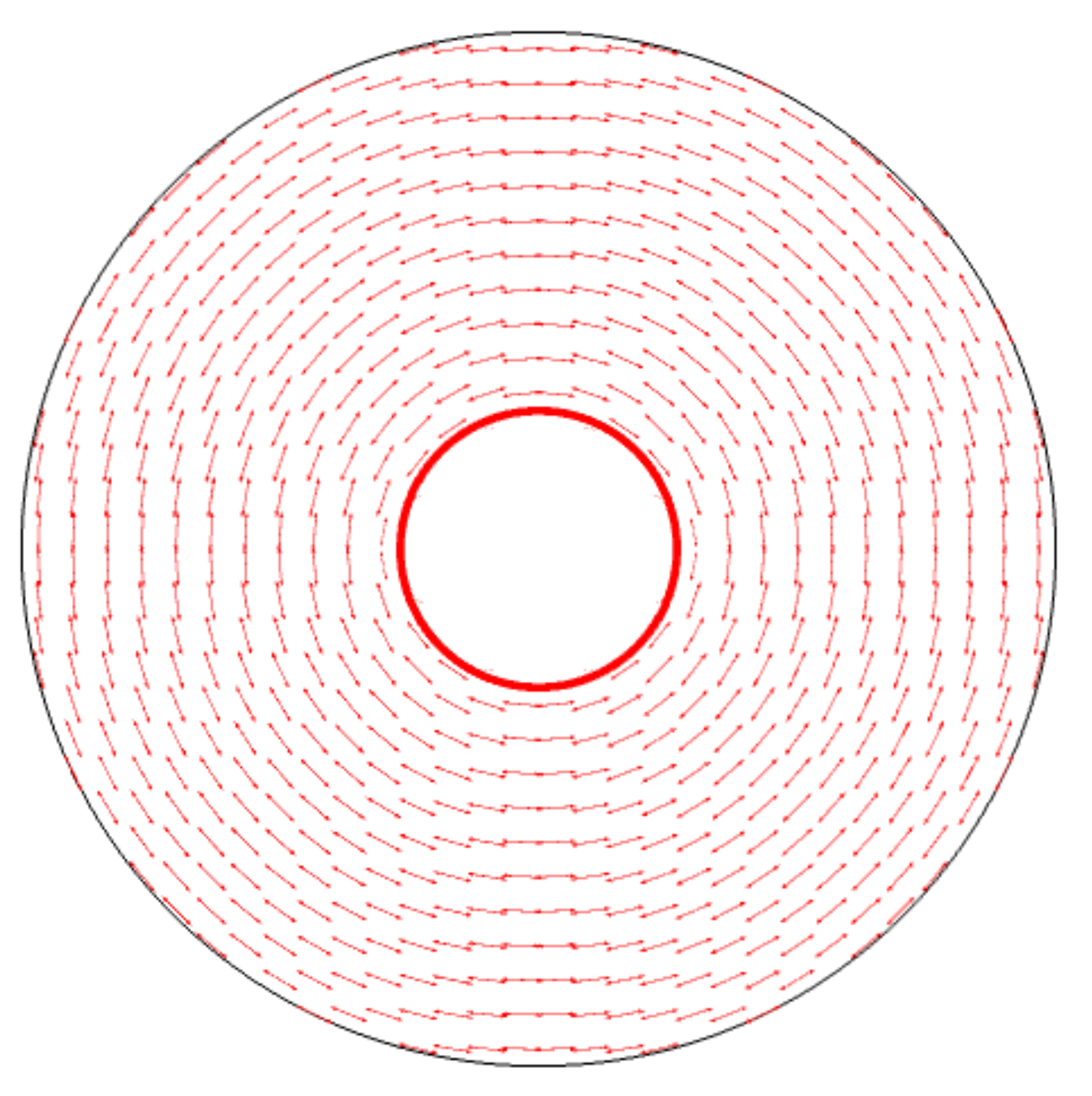}
        
        \vspace{3mm}
        
        \includegraphics[width=.4\linewidth,height=.4\linewidth]{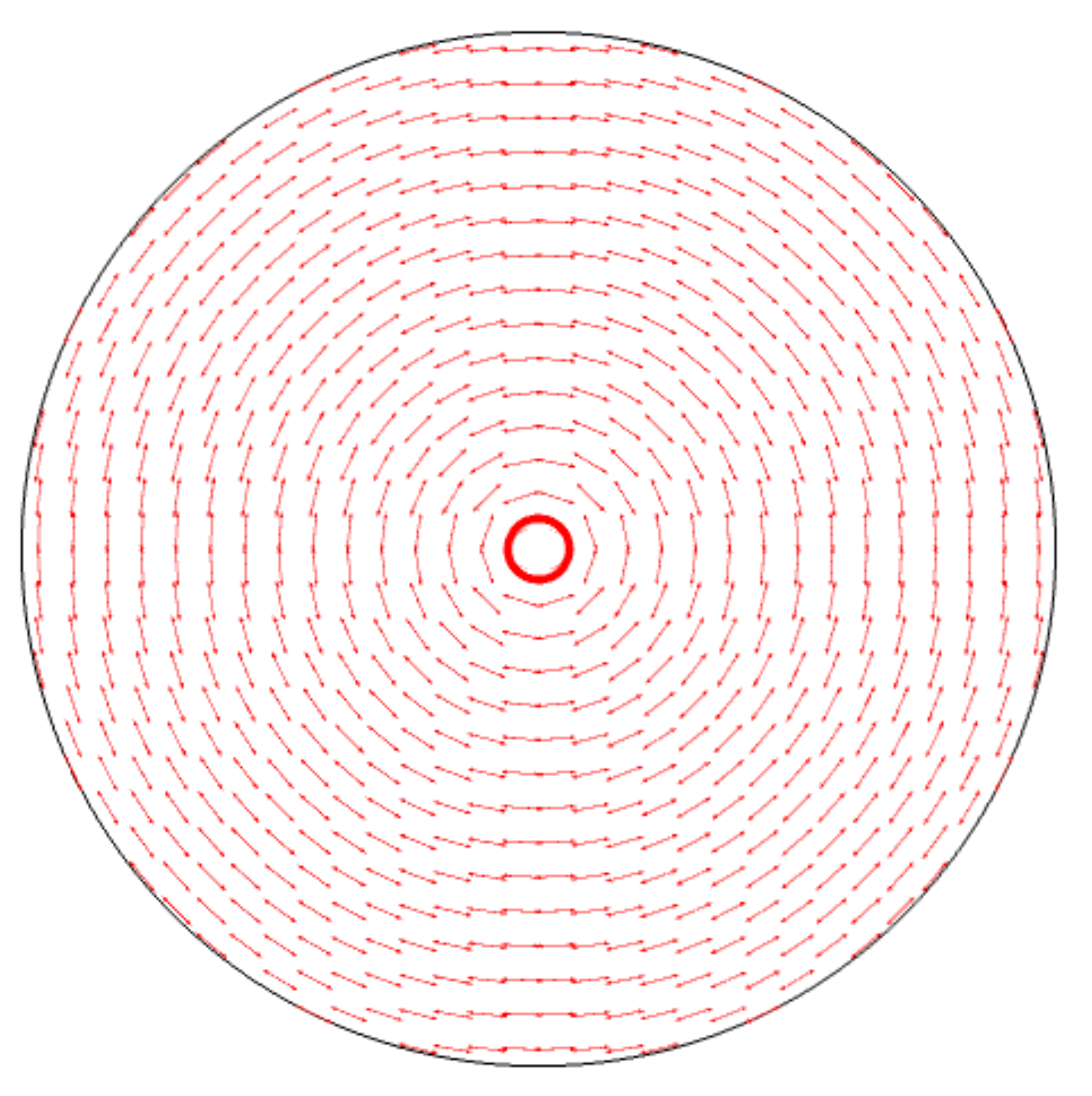} \quad
        \includegraphics[width=.4\linewidth,height=.4\linewidth]{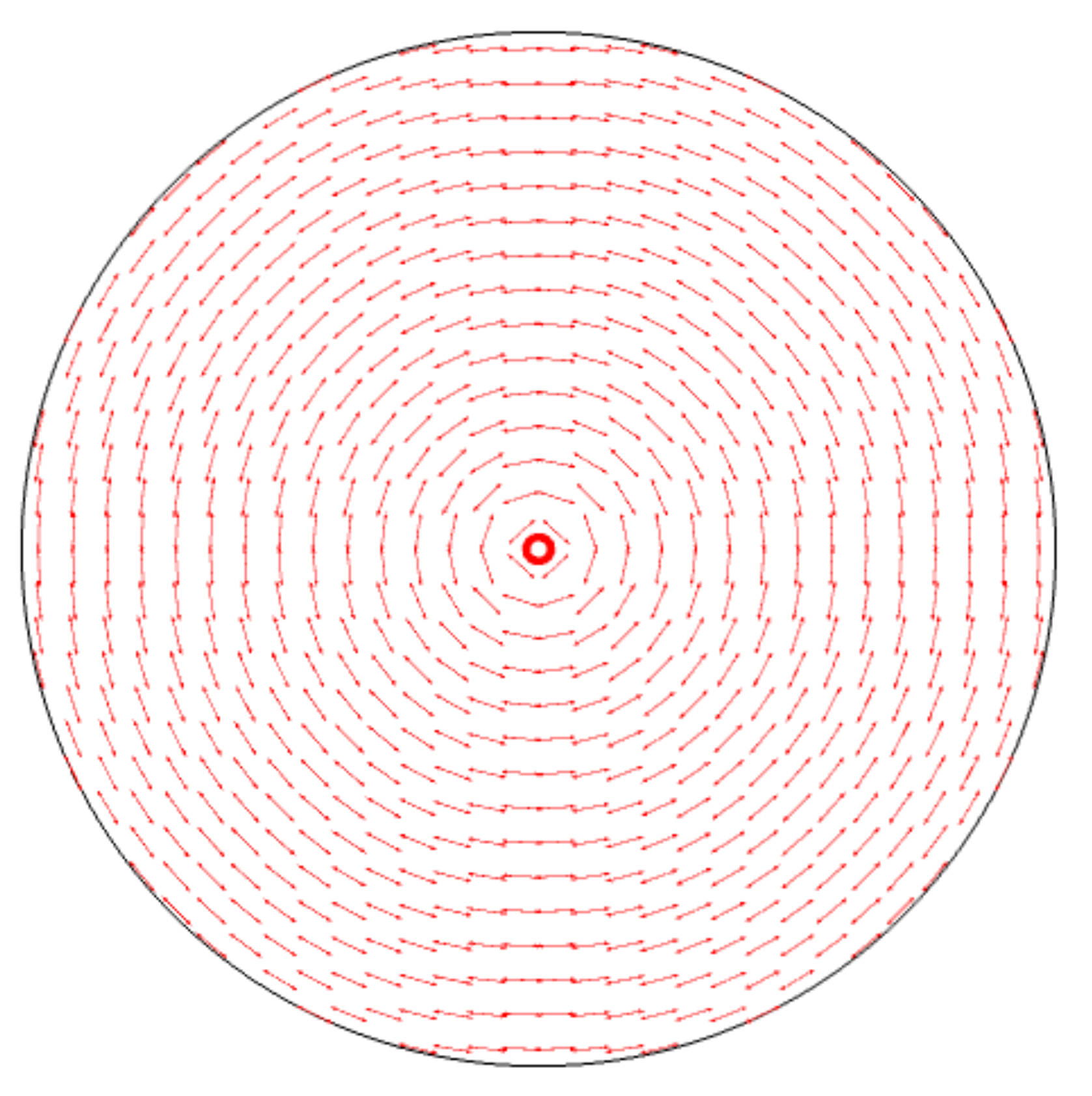}
        
        \vspace{3mm}
        
        \includegraphics[width=.4\linewidth,height=.4\linewidth]{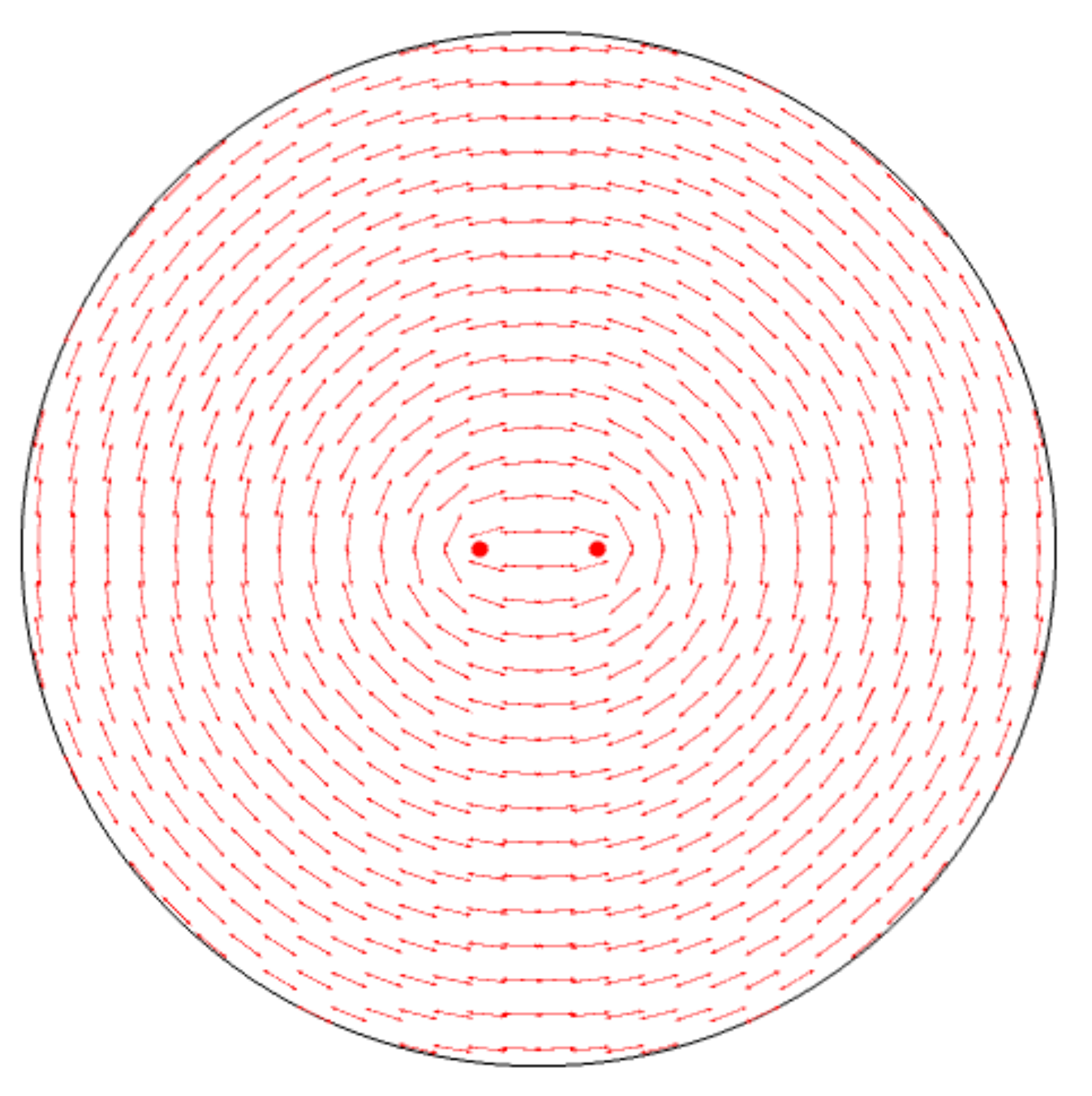}\quad \includegraphics[width=.4\linewidth,height=.4\linewidth]{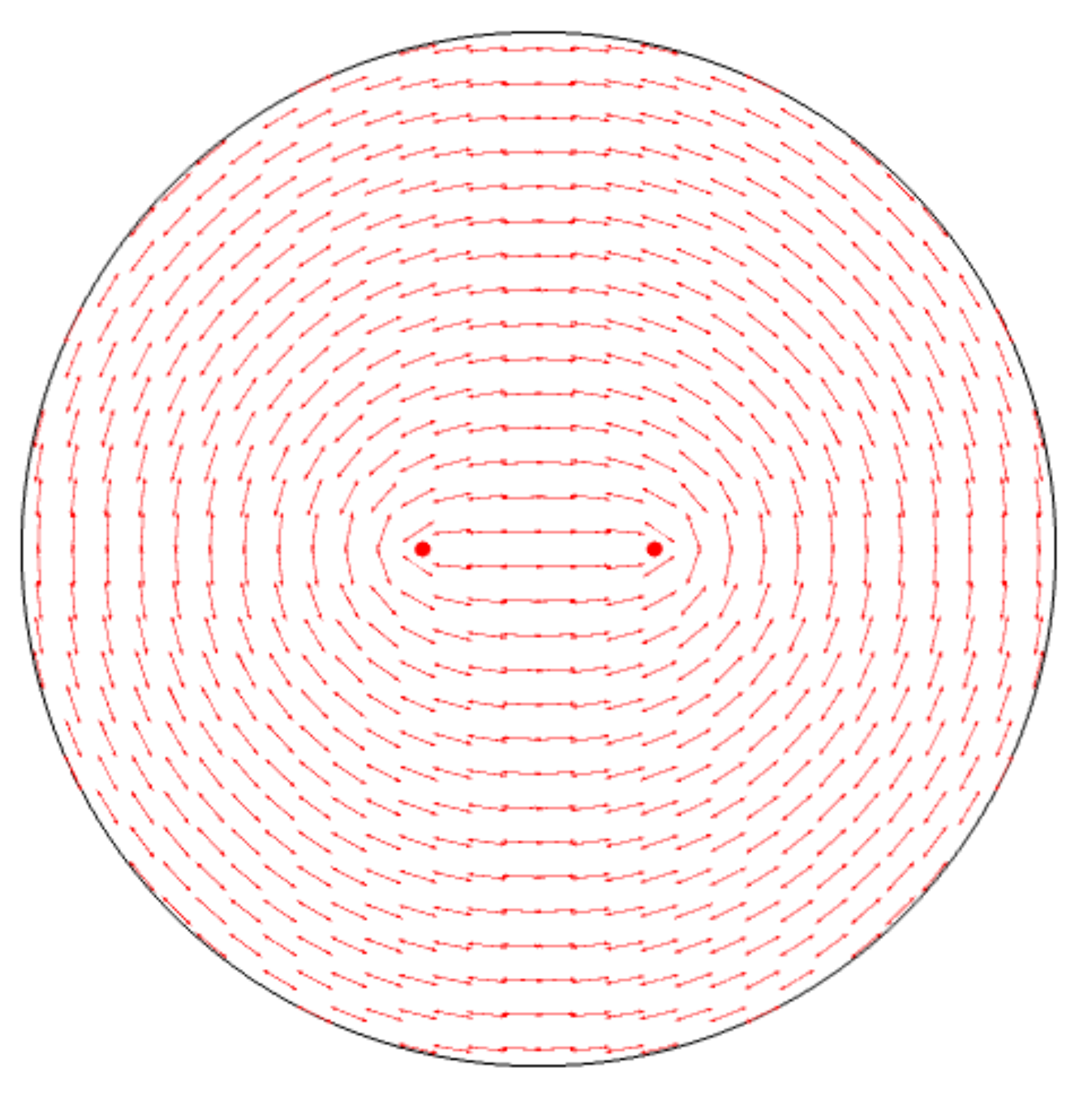}
        \caption{Simulated evolution of a degree $1$ tactoid. Here the director field is set to be equal to $(-\sin{\theta},\cos{\theta})$ on the boundary of the disk and $\theta$ is a polar angle. (cf. Fig.\ref{fig:your_label}, right tactoid). The thick red line indicates the position of the interface.} 
        \label{fig:his_label}
    \end{figure}

    The evolution of the nematic/isotropic interface is shown in Fig. \ref{fig:his_label}.
    While the isotropic tactoid is not too small, it maintains its circular shape and the director in the nematic region remains parallel to the boundary. Indeed, in this configuration the divergence contribution vanishes, the director is parallel to the interface, and the perimeter of the isotropic tactoid is minimal given its shape. 
    
    When the tactoid eventually shrinks to a vortex size seen in the right inset in the second row in Fig. \ref{fig:his_label}, it loses stability and splits into two degree $1/2$ vortices that drift away from each other with time. The overall behavior is qualitatively similar to that of the right tactoid in Fig. \ref{fig:your_label}.
    
    \subsection{Degree $-1$ tactoid}

\begin{figure}[htp]
        \centering
        \includegraphics[width=.4\linewidth,height=.4\linewidth]{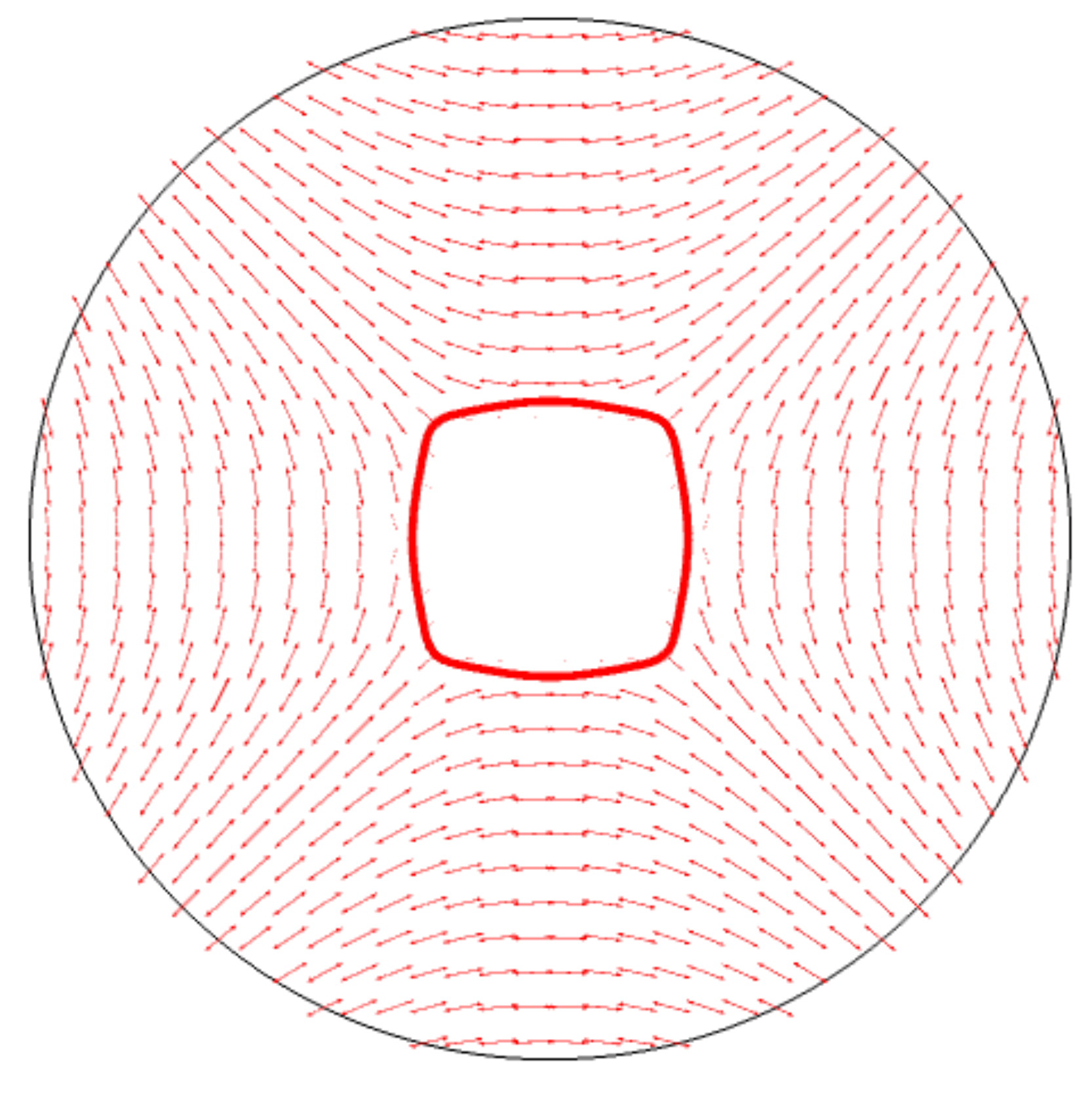}\quad
        \includegraphics[width=.4\linewidth,height=.4\linewidth]{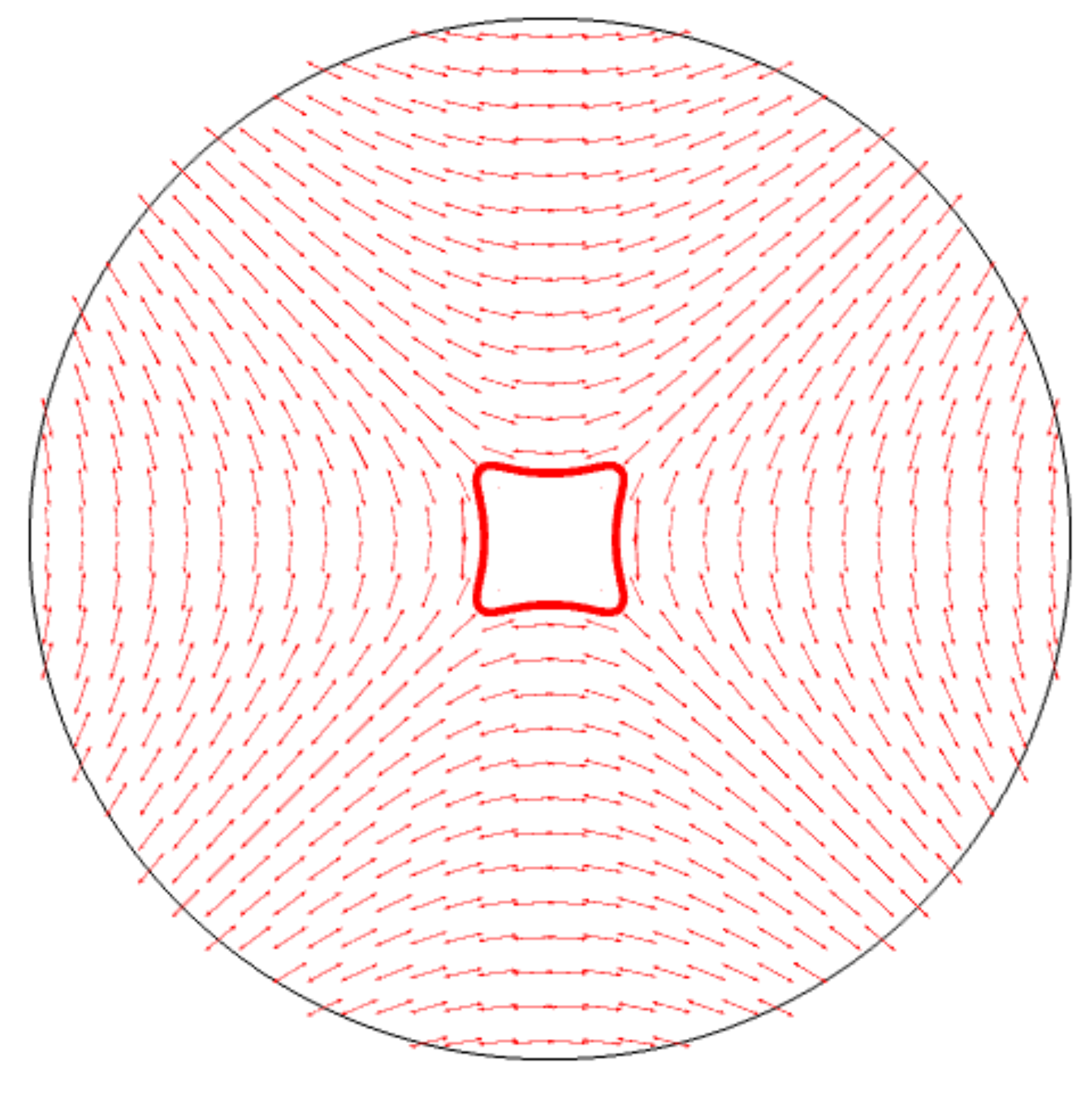}
        
        \vspace{3mm}
        
        \includegraphics[width=.4\linewidth,height=.4\linewidth]{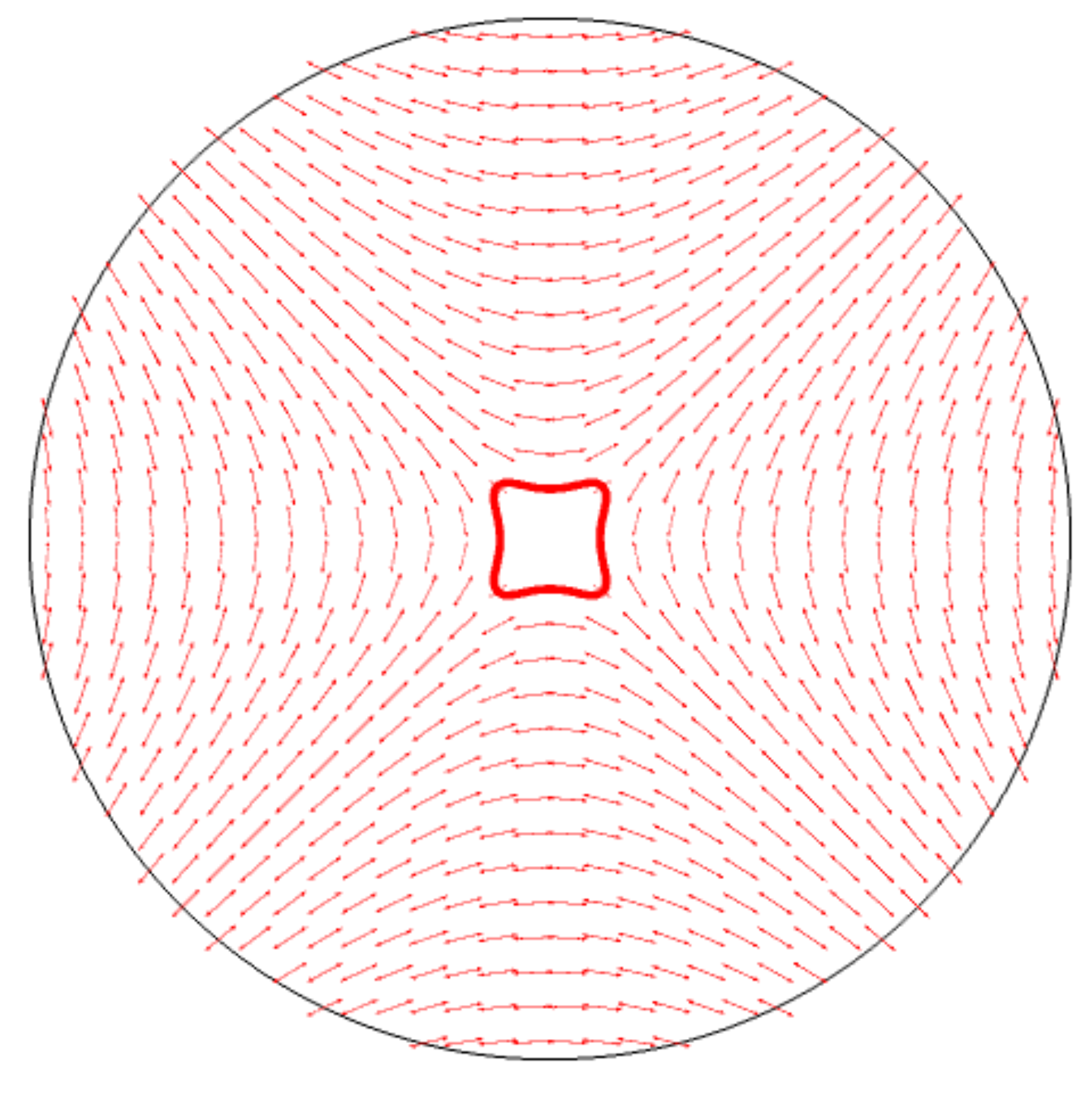}\quad \includegraphics[width=.4\linewidth,height=.4\linewidth]{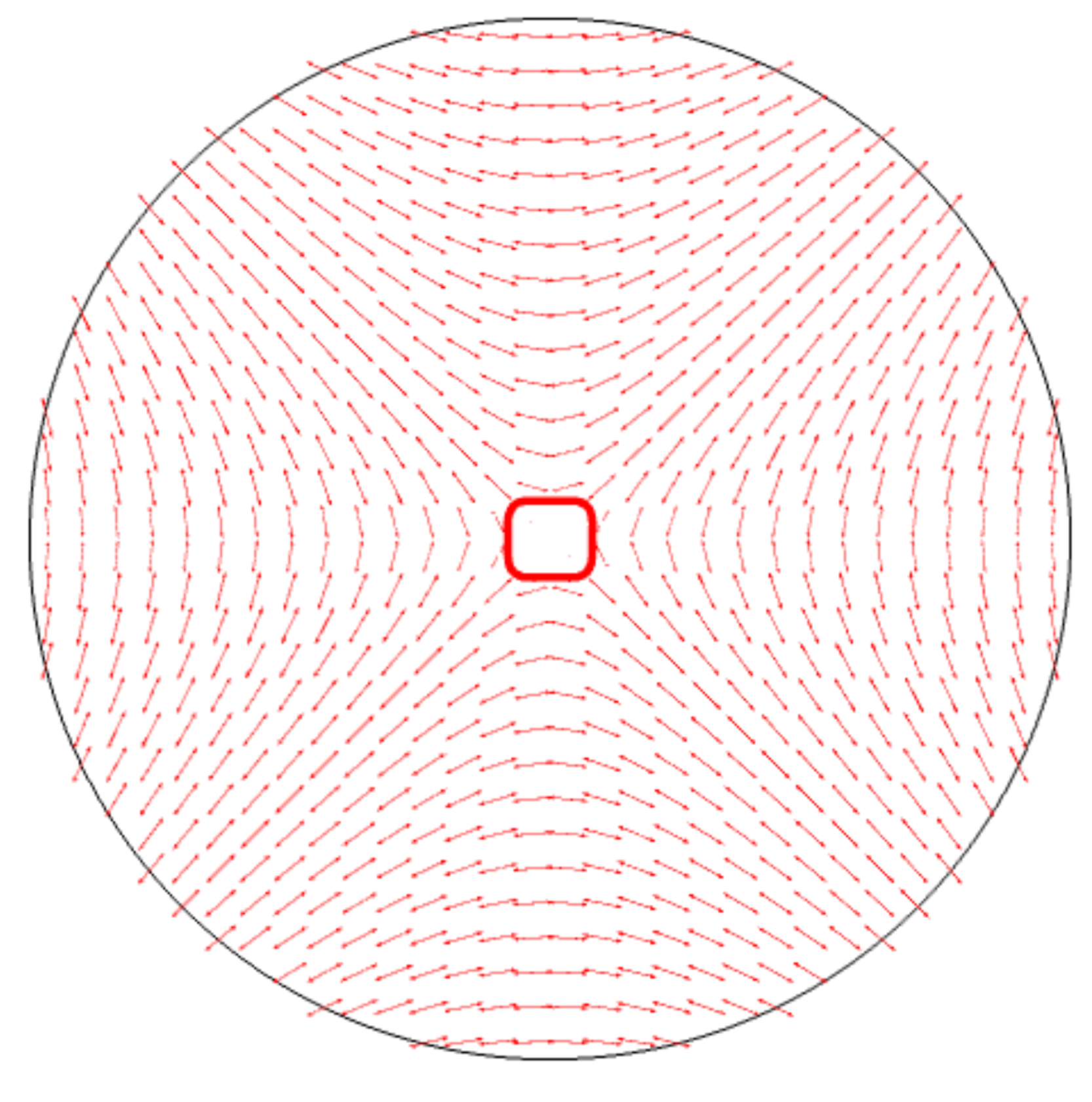}
        
        \vspace{3mm}
        
        \includegraphics[width=.4\linewidth,height=.4\linewidth]{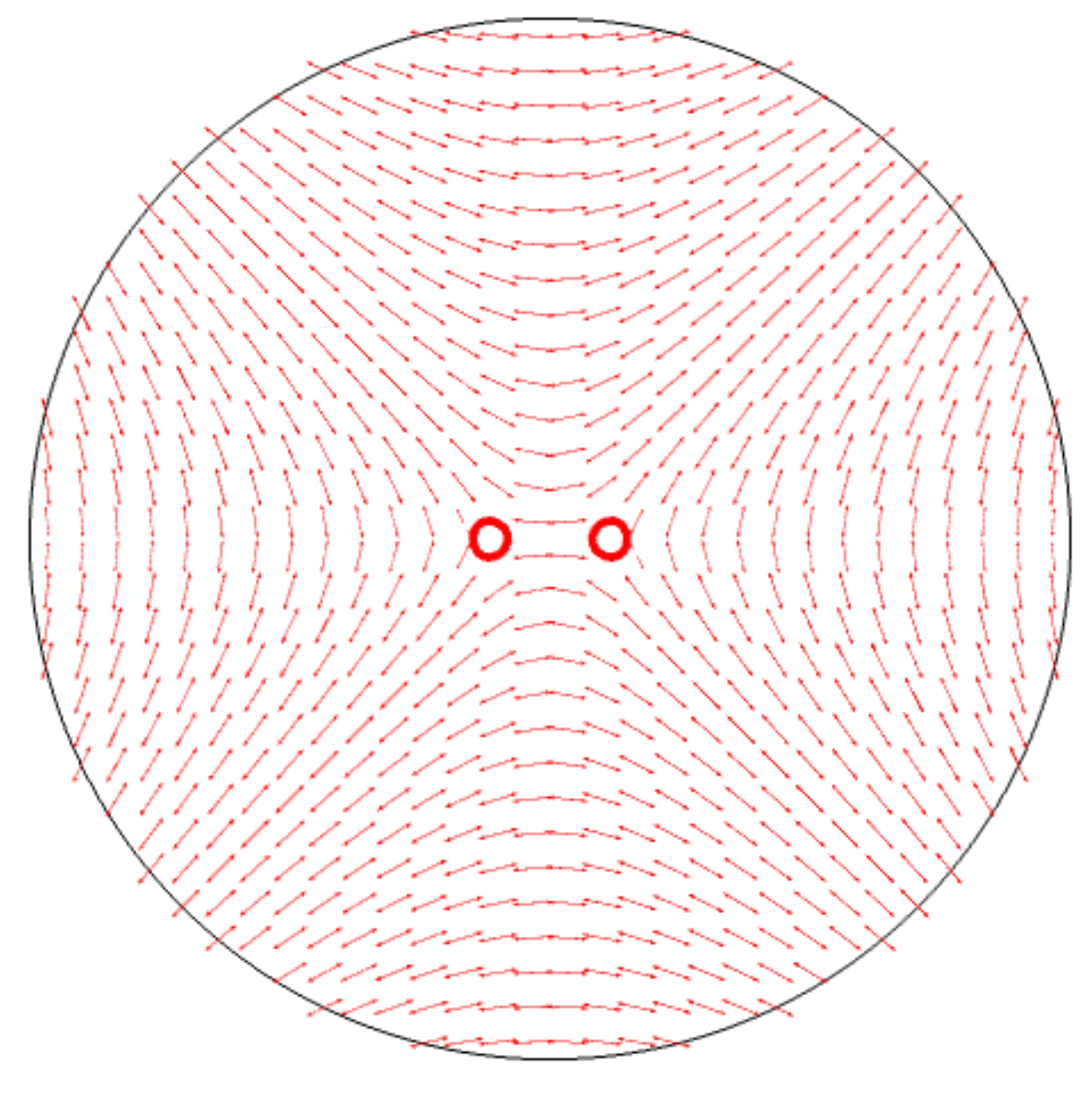}\quad \includegraphics[width=.4\linewidth,height=.4\linewidth]{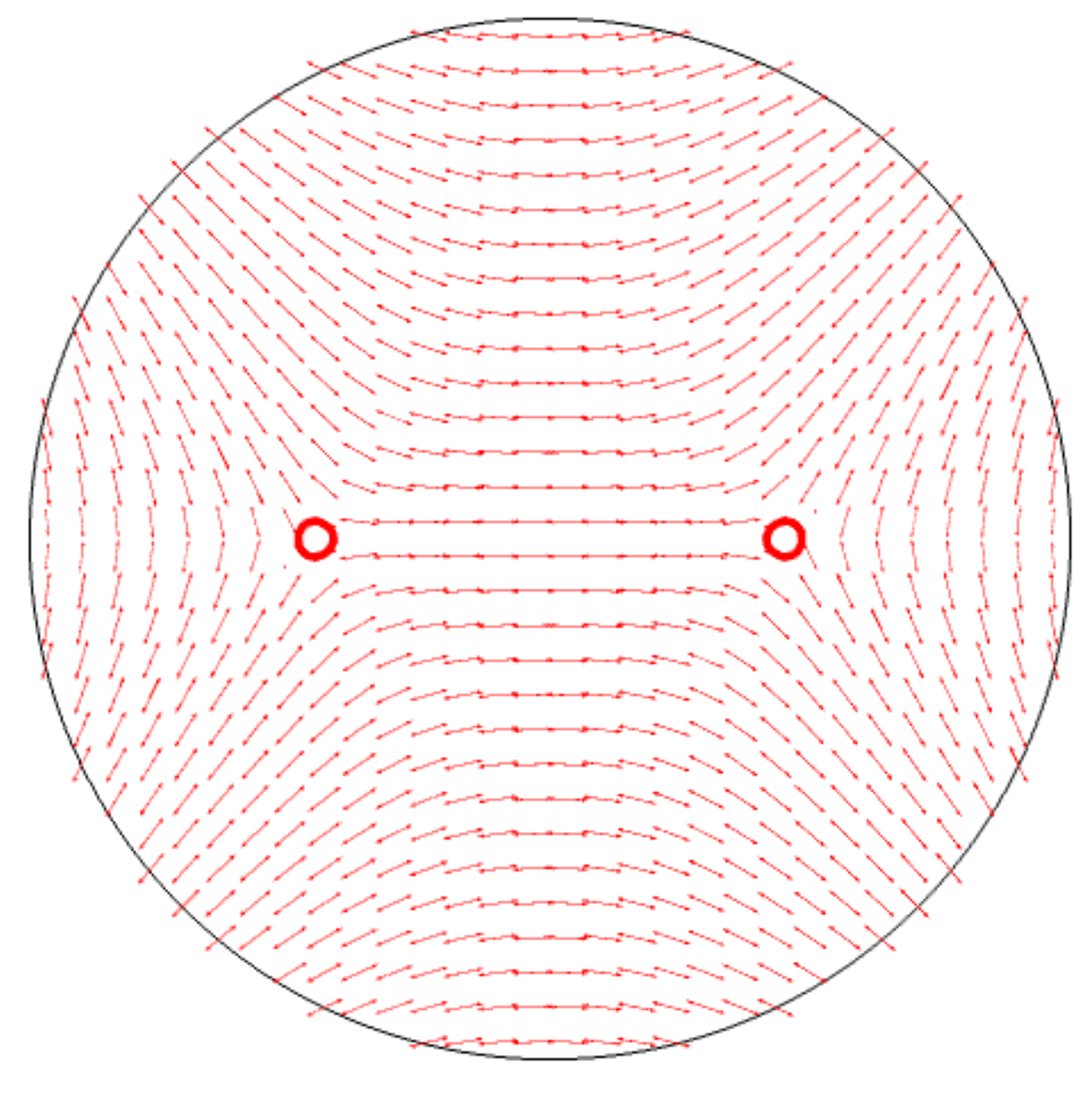}
        \caption{Simulated evolution of a degree $-1$ tactoid. Here the director field is set to be equal to $(-\cos{\theta},\sin{\theta})$ on the boundary of the disk and $\theta$ is a polar angle. (cf. Fig.\ref{fig:my_label}). The thick red line indicates the position of the interface.}
        \label{fig:hers_label}
    \end{figure}

    We now return to the boundary conditions
    \[\beta|_{\partial\omega}=\frac{1}{2},\quad u|_{\partial\omega}=\frac{3}{2}(-\cos{2\theta},\sin{2\theta}),\]
    considered in the beginning of this section for zero undercooling. Here we assume instead that
    $\alpha=-0.1$ and look for the effects on evolution of the bias between the values of local minima of the potential energy corresponding to the nematic and isotropic state. This bias is associated with lowering the temperature below that of the nematic-to-isotropic transition. The corresponding numerical results are shown in Fig. \ref{fig:hers_label}. 
    The first three figures essentially demonstrate the development of a configuration shown in Fig. \ref{fig:his_label}. The larger thermodynamic forces driving the phase transition in the present case, however, push the size of the tactoid further down essentially to that of a vortex core. At this point, the isotropic region loses stability and splits into two vortices of degree $-1/2$, similar to what can be seen in Fig. \ref{fig:my_label}.
    
  \subsection{Degree $0$ tactoid}

\begin{figure}[htp]
        \centering
        \includegraphics[width=.4\linewidth,height=.4\linewidth]{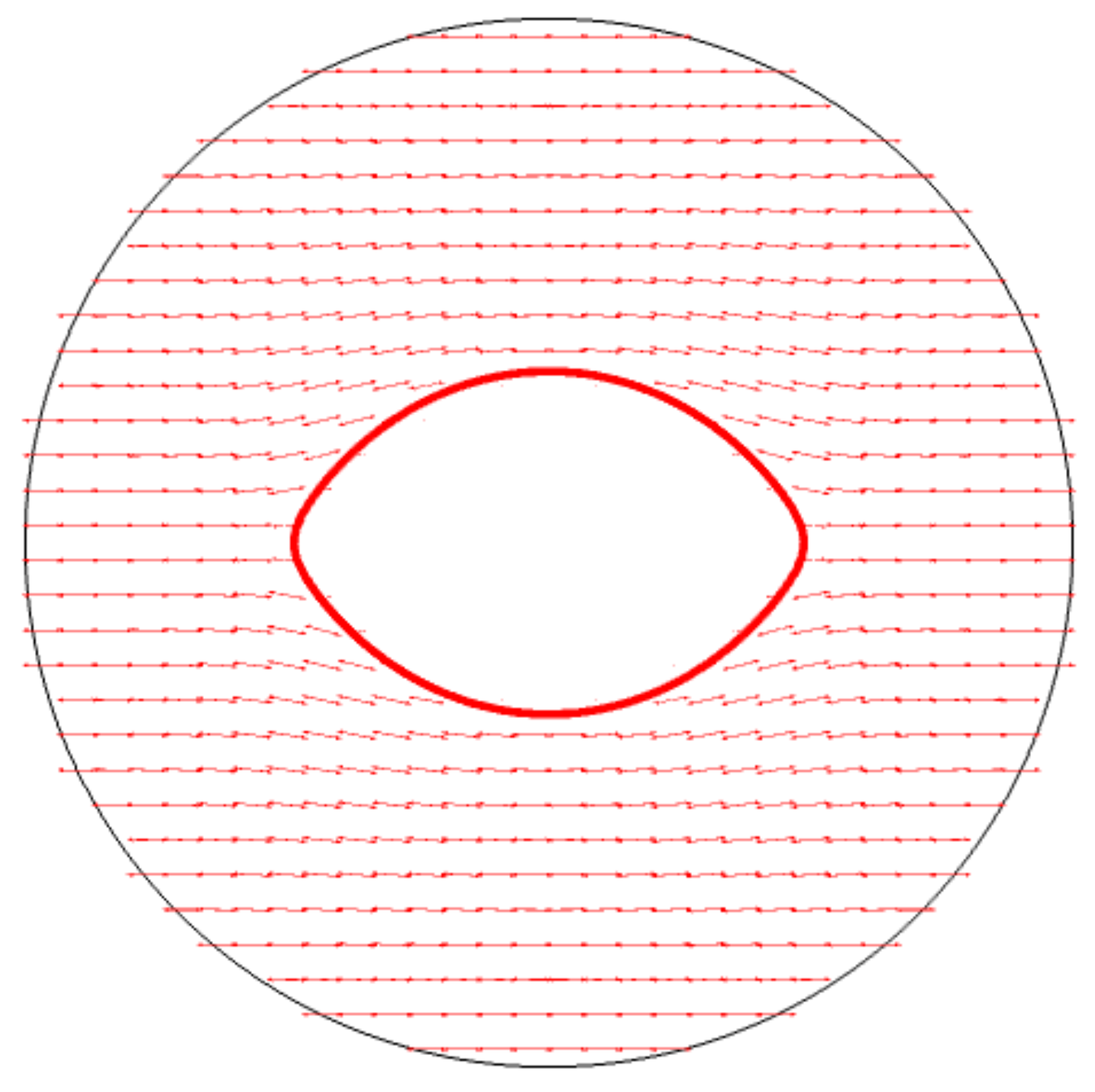}\quad
        \includegraphics[width=.4\linewidth,height=.4\linewidth]{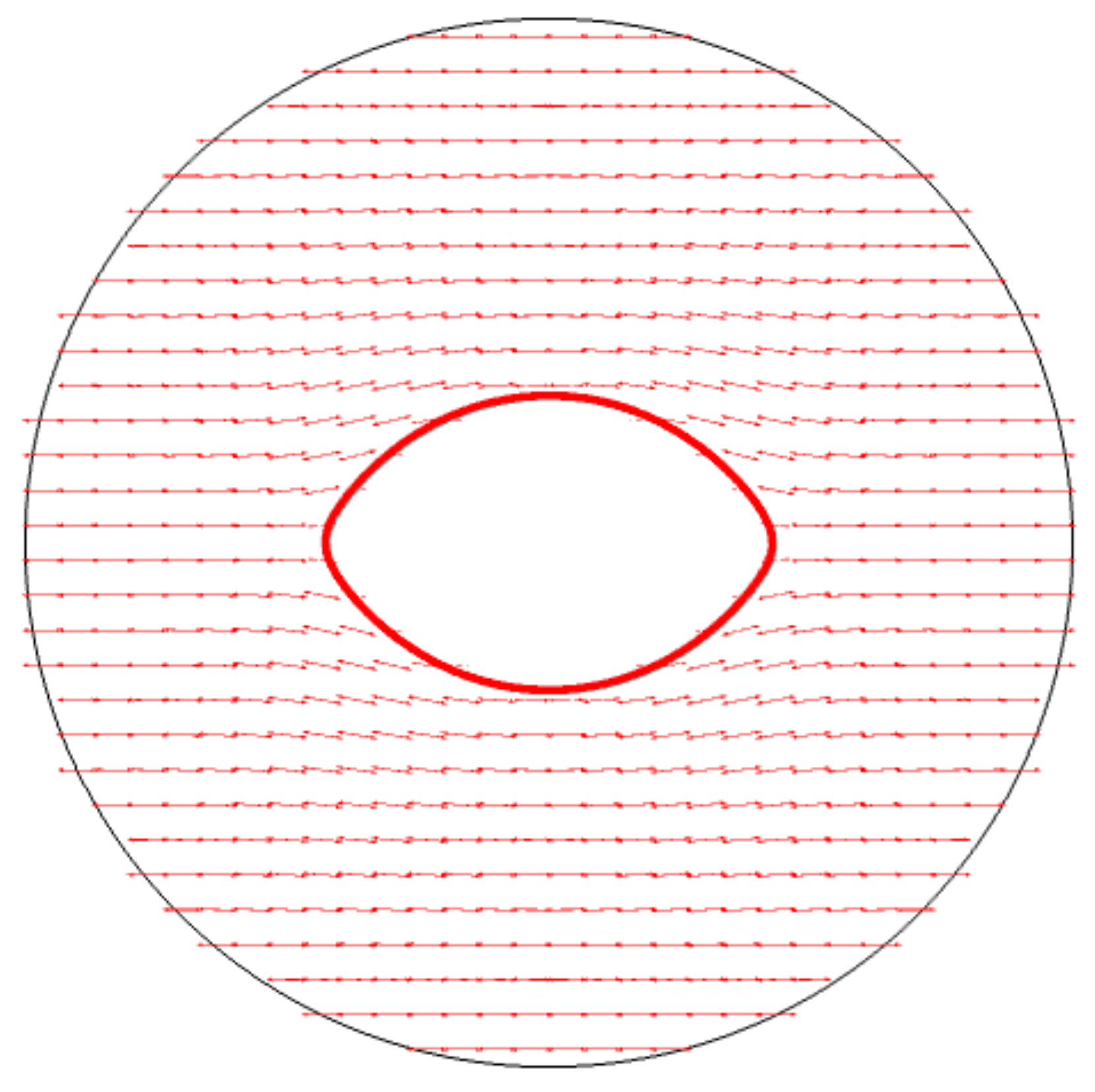}
        
        \vspace{3mm}
        
        \includegraphics[width=.4\linewidth,height=.4\linewidth]{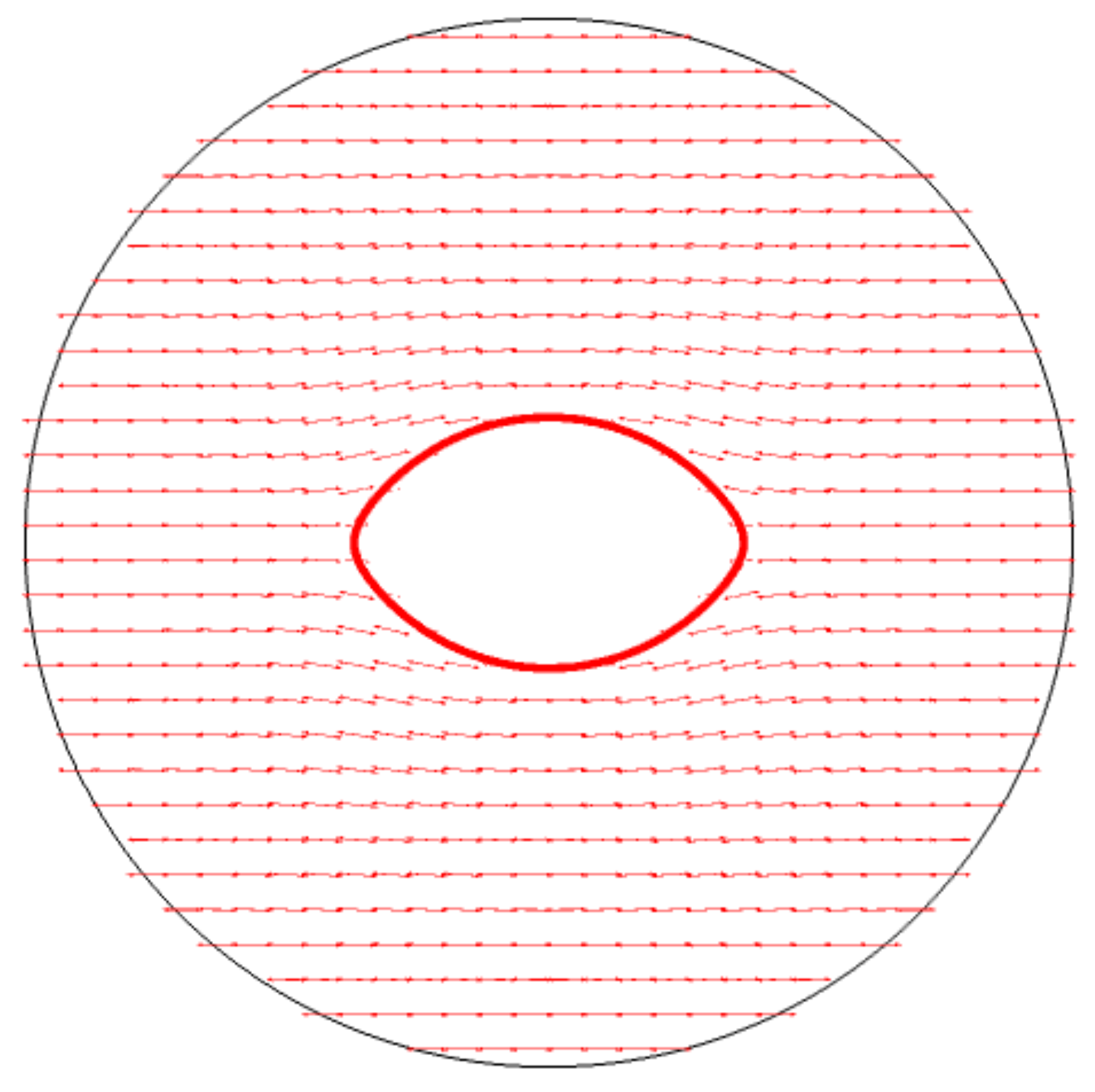}\quad \includegraphics[width=.4\linewidth,height=.4\linewidth]{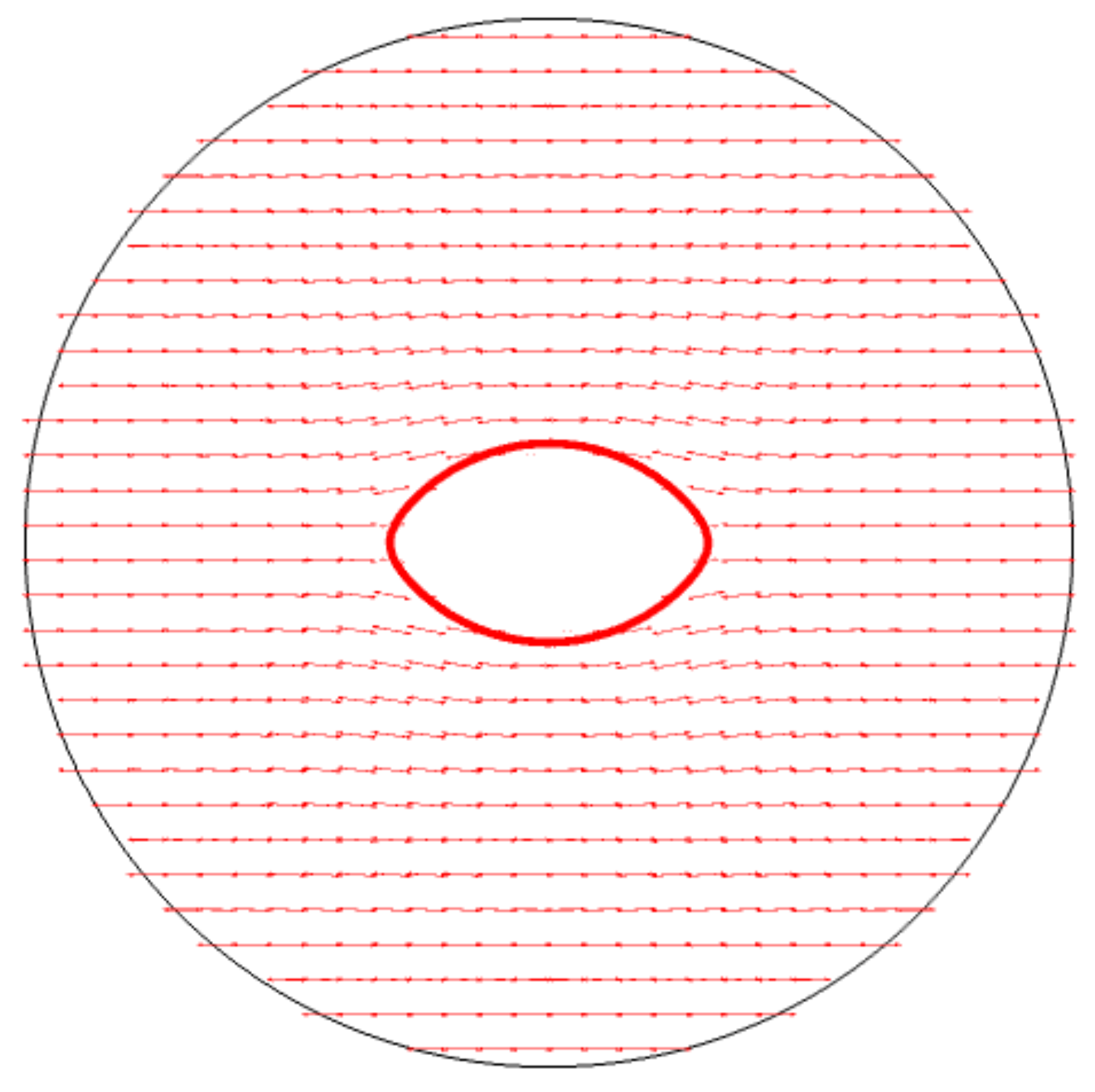}
        
        \vspace{3mm}
        
        \includegraphics[width=.4\linewidth,height=.4\linewidth]{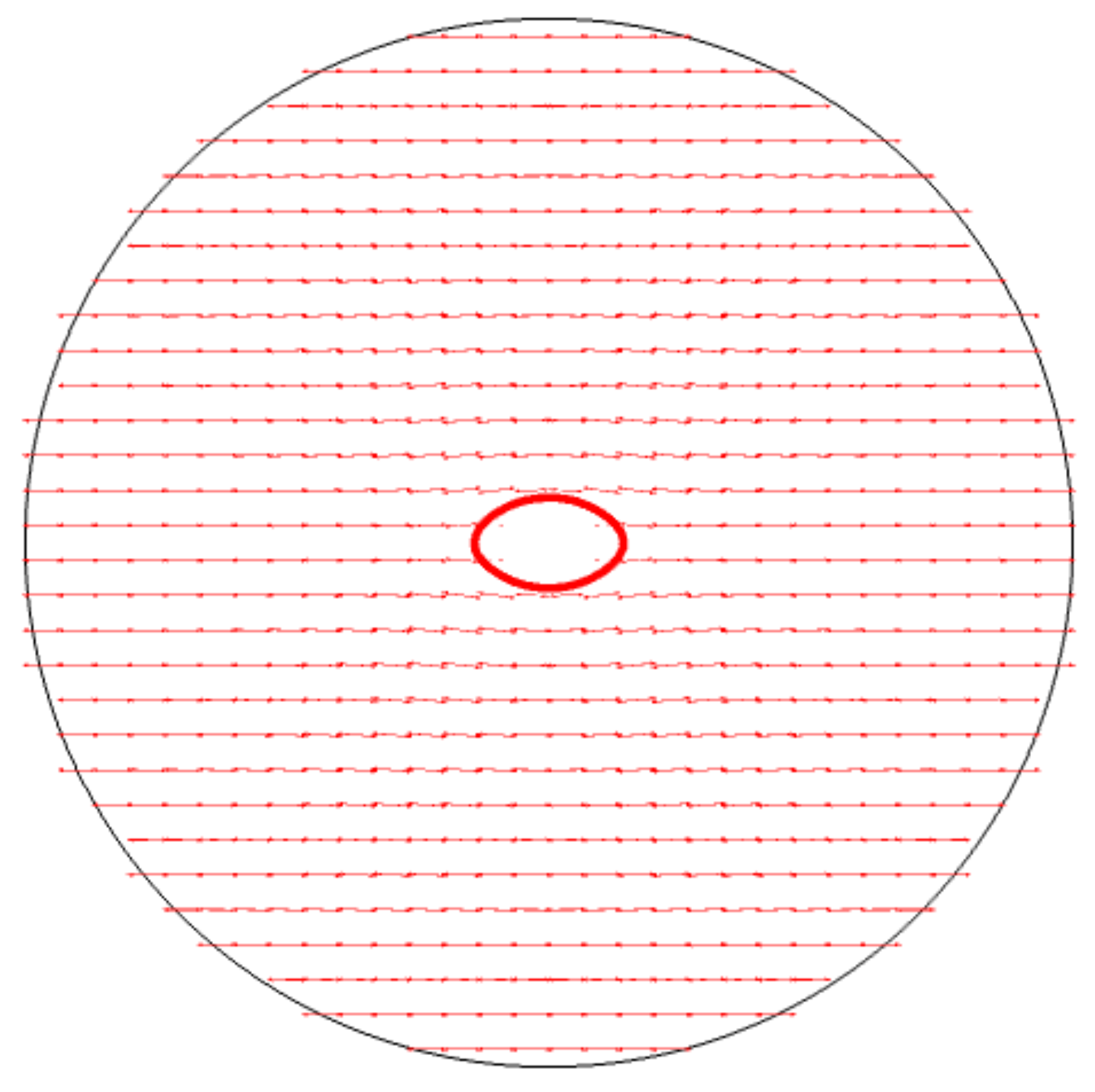}\quad \includegraphics[width=.4\linewidth,height=.4\linewidth]{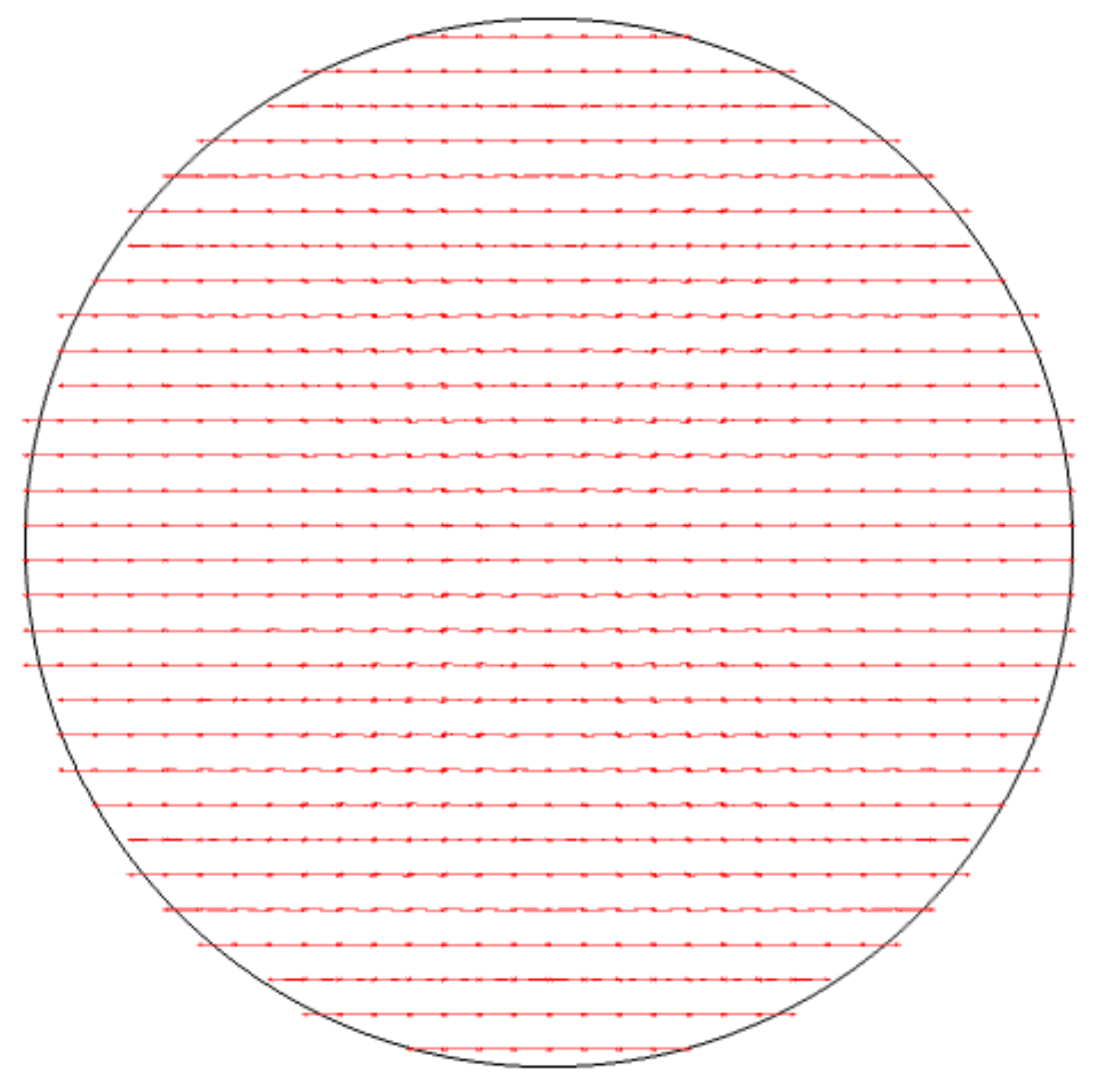}
        \caption{Simulated evolution of a degree $0$ tactoid. Here the director field is set to be equal to $(1,0)$ on the boundary of the disk. (cf. Fig.\ref{fig:your_label}, left tactoid). The thick red line indicates the position of the interface.}
        \label{fig:their_label}
    \end{figure}

    Next, we impose the constant boundary conditions
    \[\beta|_{\partial\omega}=\frac{1}{2},\quad u|_{\partial\omega}=\frac{3}{2}(1,0),\]
    and suppose that $\alpha=0$. The numerically computed evolution of a degree zero tactoid that results is shown in Fig. \ref{fig:their_label} and qualitatively resembles the behavior of a similar tactoid in the experiment as depicted in Fig.  \ref{fig:your_label}.
    A similar shape is also seen in evolution of degree zero interfaces in the CSH-director model in \cite{GNSV} and is explained by the fact that the director has to be parallel to the interface. In particular, the interface cannot be smooth, for if it were, then it would carry a nonzero topological degree different from the degree on the boundary of the domain.
    \begin{figure}[htp]
        \centering
        \includegraphics[width=.4\linewidth,height=.4\linewidth]{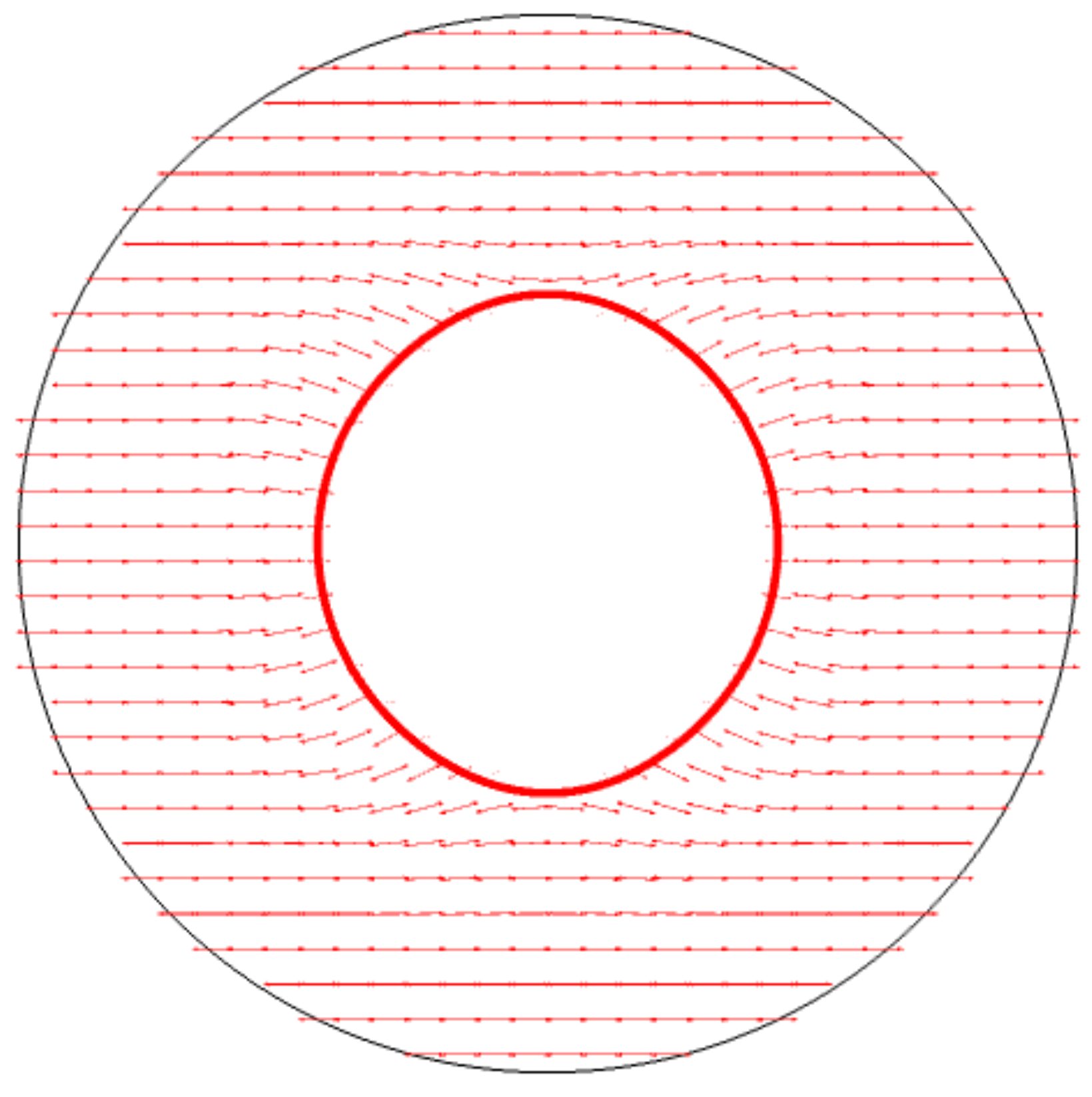}\quad
        \includegraphics[width=.4\linewidth,height=.4\linewidth]{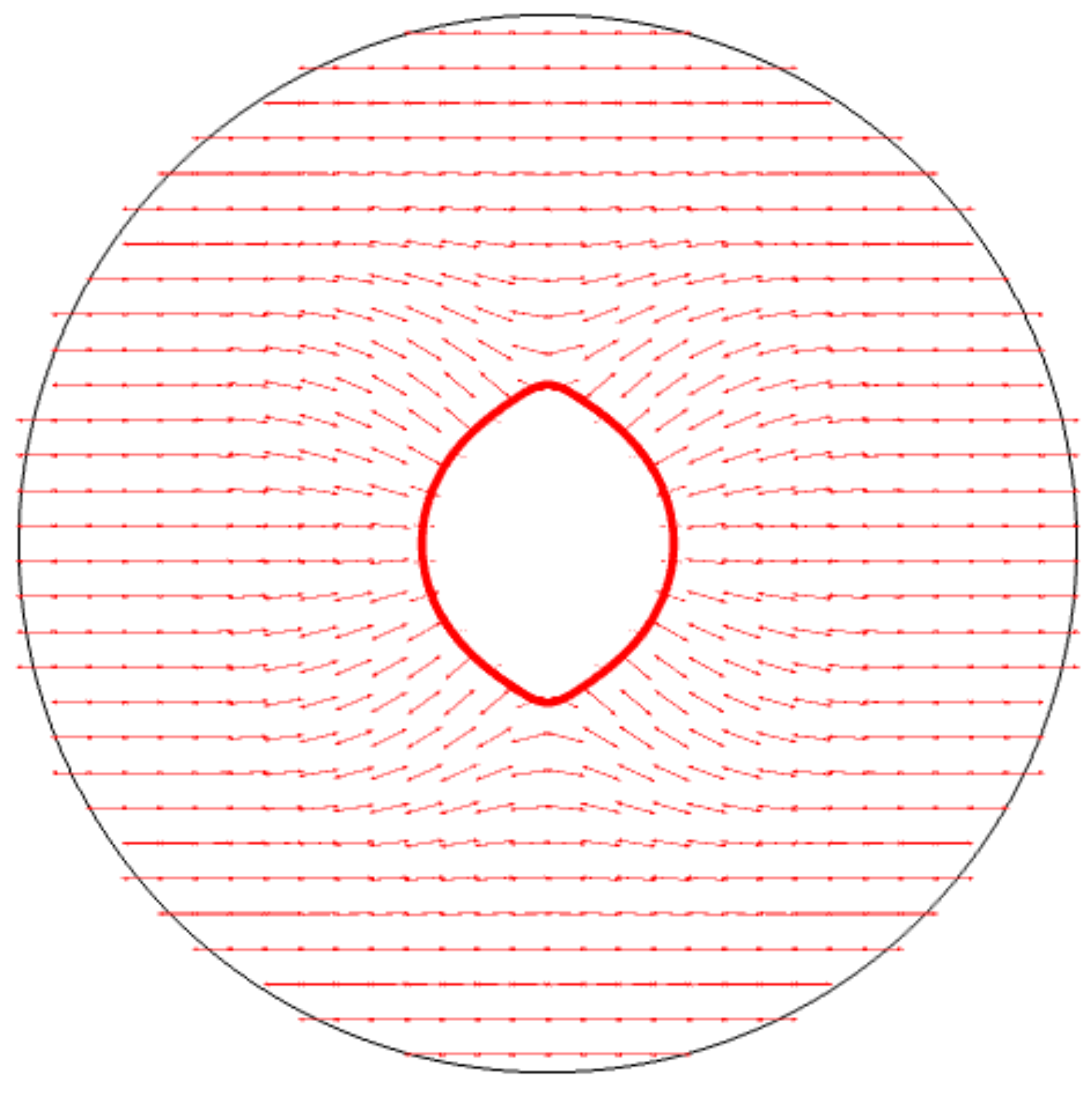}
        
        \vspace{3mm}
        
        \includegraphics[width=.4\linewidth,height=.4\linewidth]{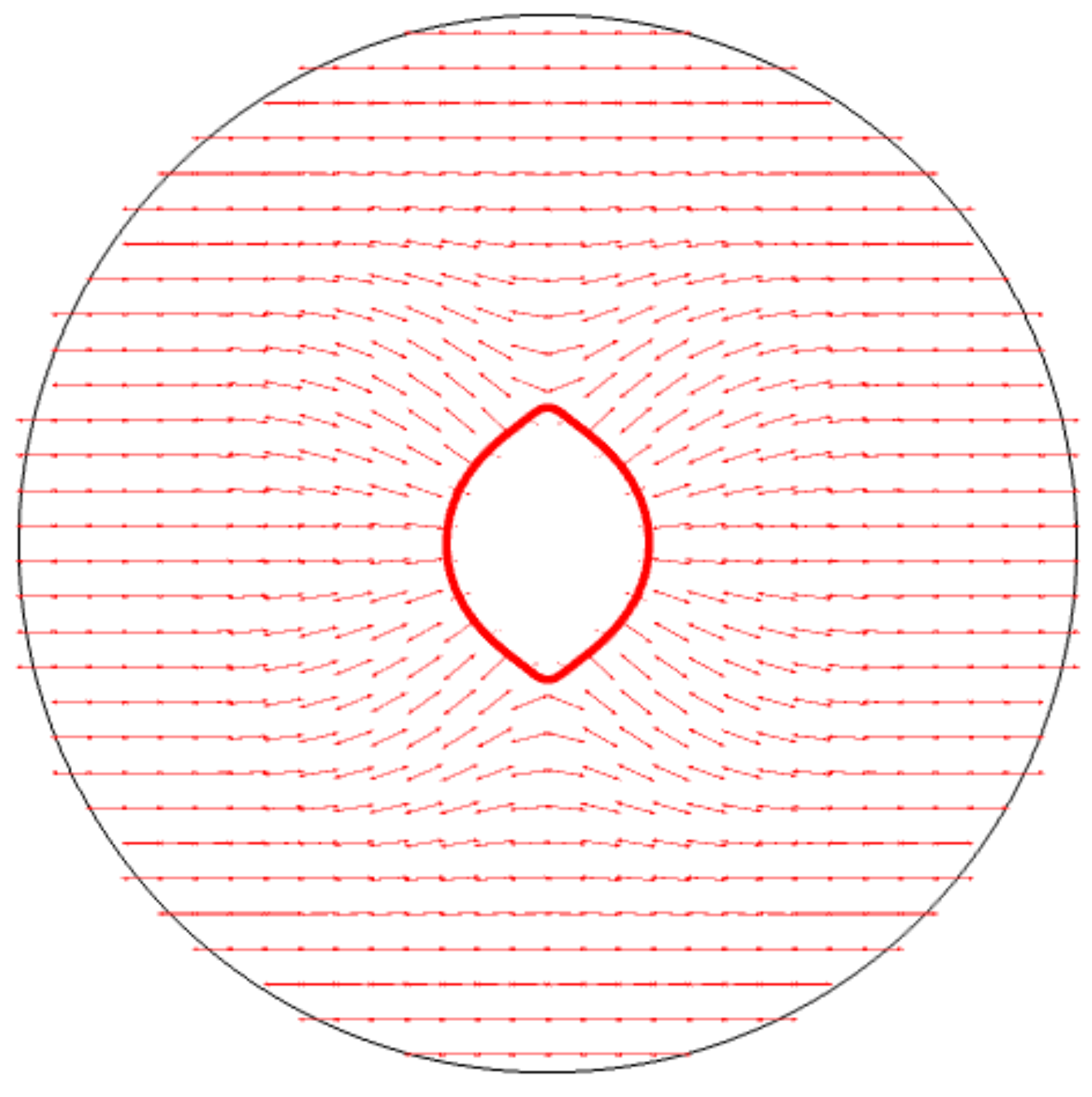}\quad \includegraphics[width=.4\linewidth,height=.4\linewidth]{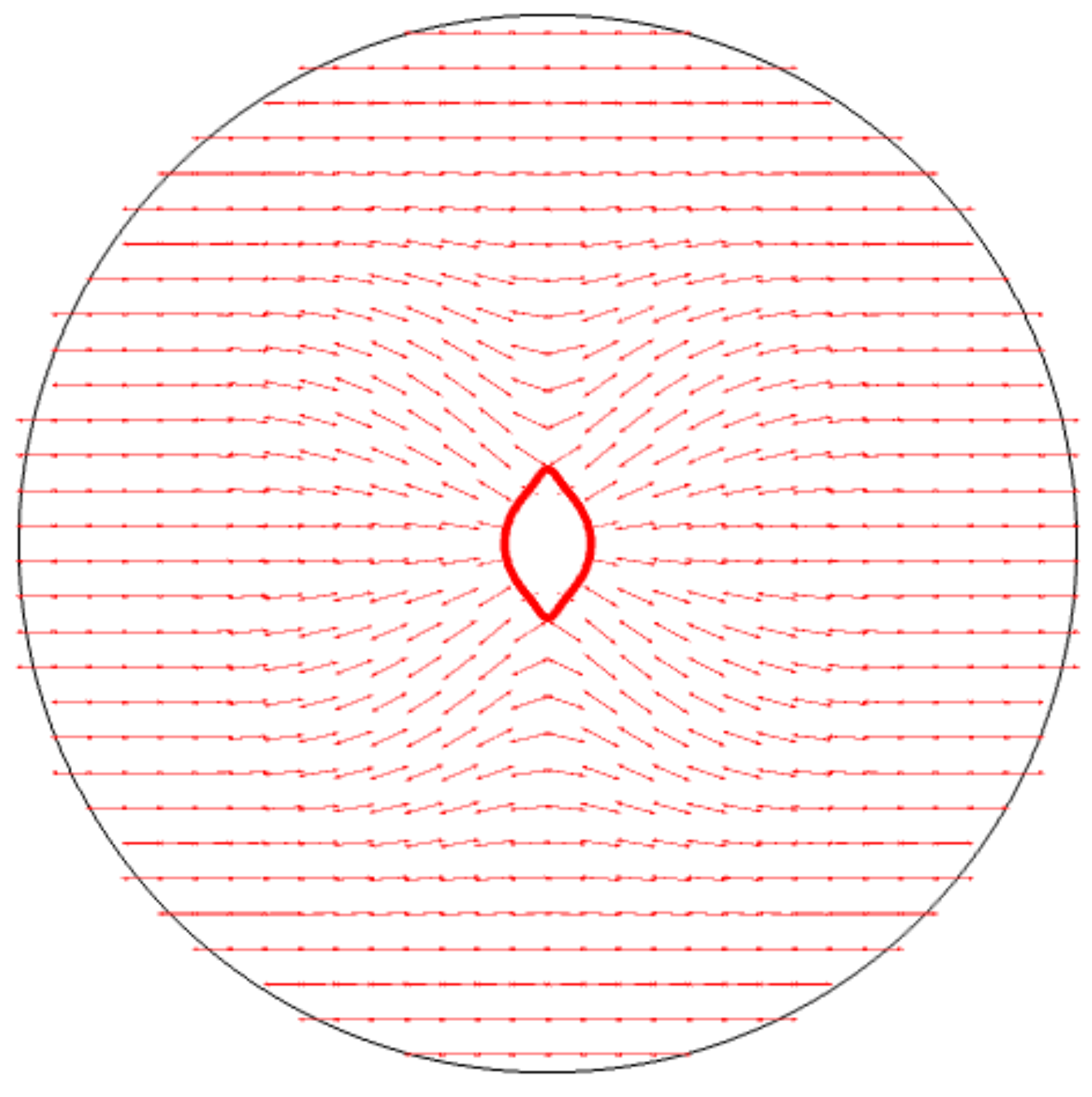}
        
        \vspace{3mm} 
        
        \includegraphics[width=.4\linewidth,height=.4\linewidth]{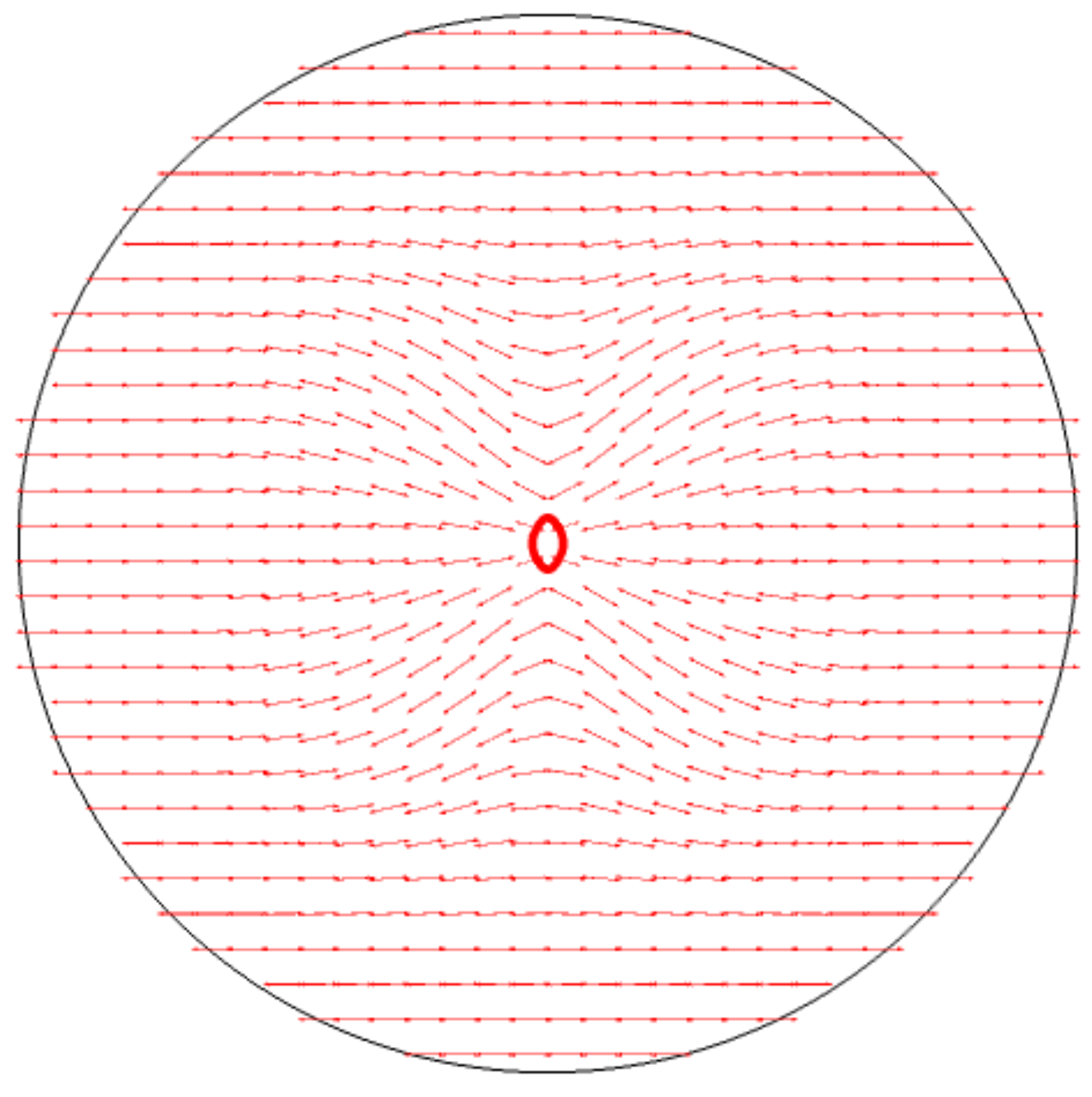}\quad \includegraphics[width=.4\linewidth,height=.4\linewidth]{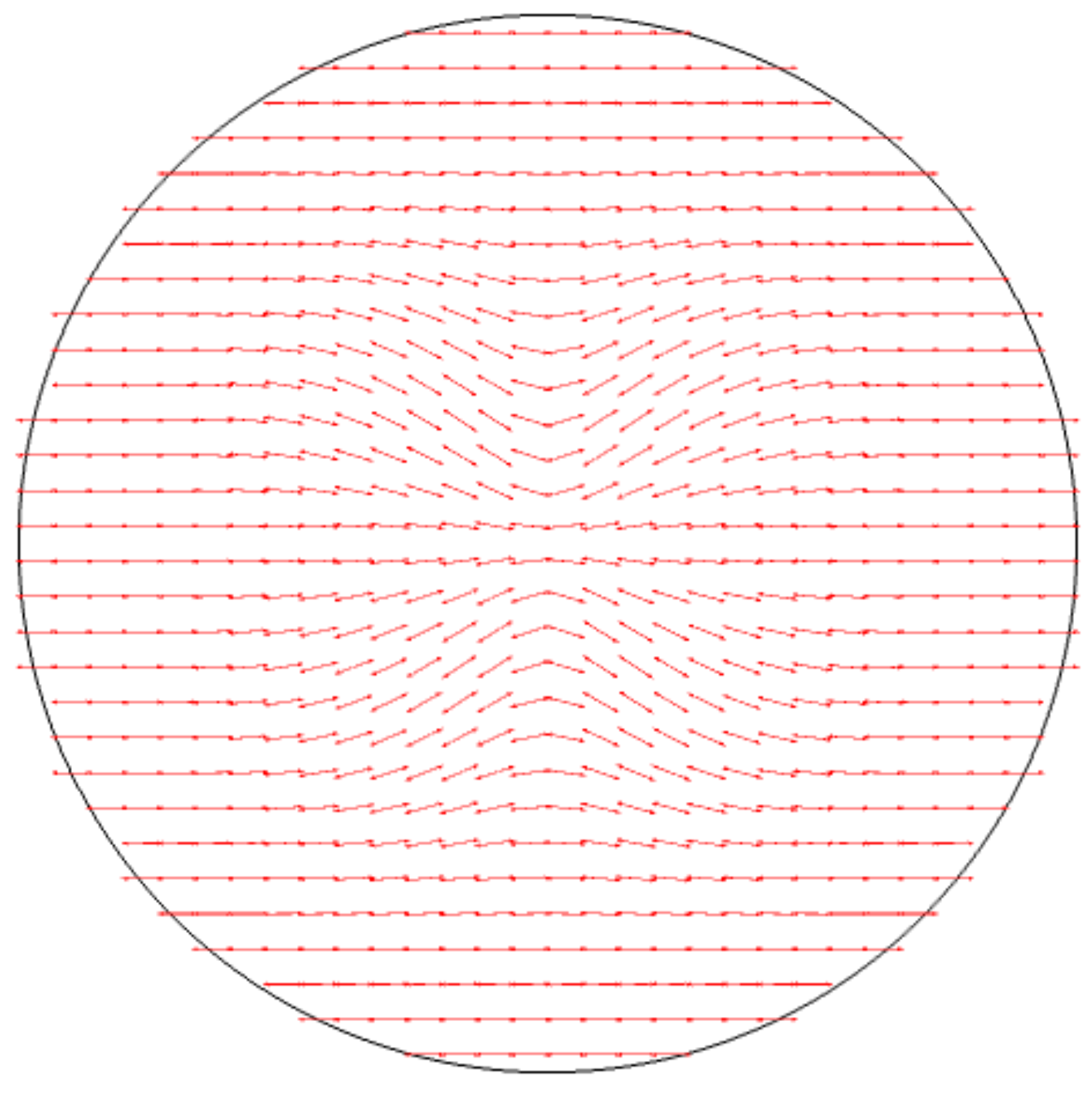}
        \caption{Simulated evolution of a degree $0$ tactoid. Here the director field is set to be equal to $(1,0)$ on the boundary of the disk and the undercooling is significantly larger than that in Fig.\ref{fig:their_label}. The thick red line indicates the position of the interface.}
        \label{fig:its_label}
    \end{figure}
    One interesting issue that we observed in the course of our simulations is illustrated in Fig. \ref{fig:its_label}. When the undercooling is large ($\alpha=0.2$), the director appears to be orthogonal to the moving interface rather than being parallel to it as would be expected. A possible explanation for this effect is that the velocity of the interface is relatively large for larger undercoolings and the mobility of the director might not be sufficient for it to relax in a proper direction. We plan to investigate this behavior further in a future work.
    
    \subsection{Coalescence of Nematic Tactoids}
    
\begin{figure}[htp]
        \centering
        \includegraphics[width=.4\linewidth,height=.4\linewidth]{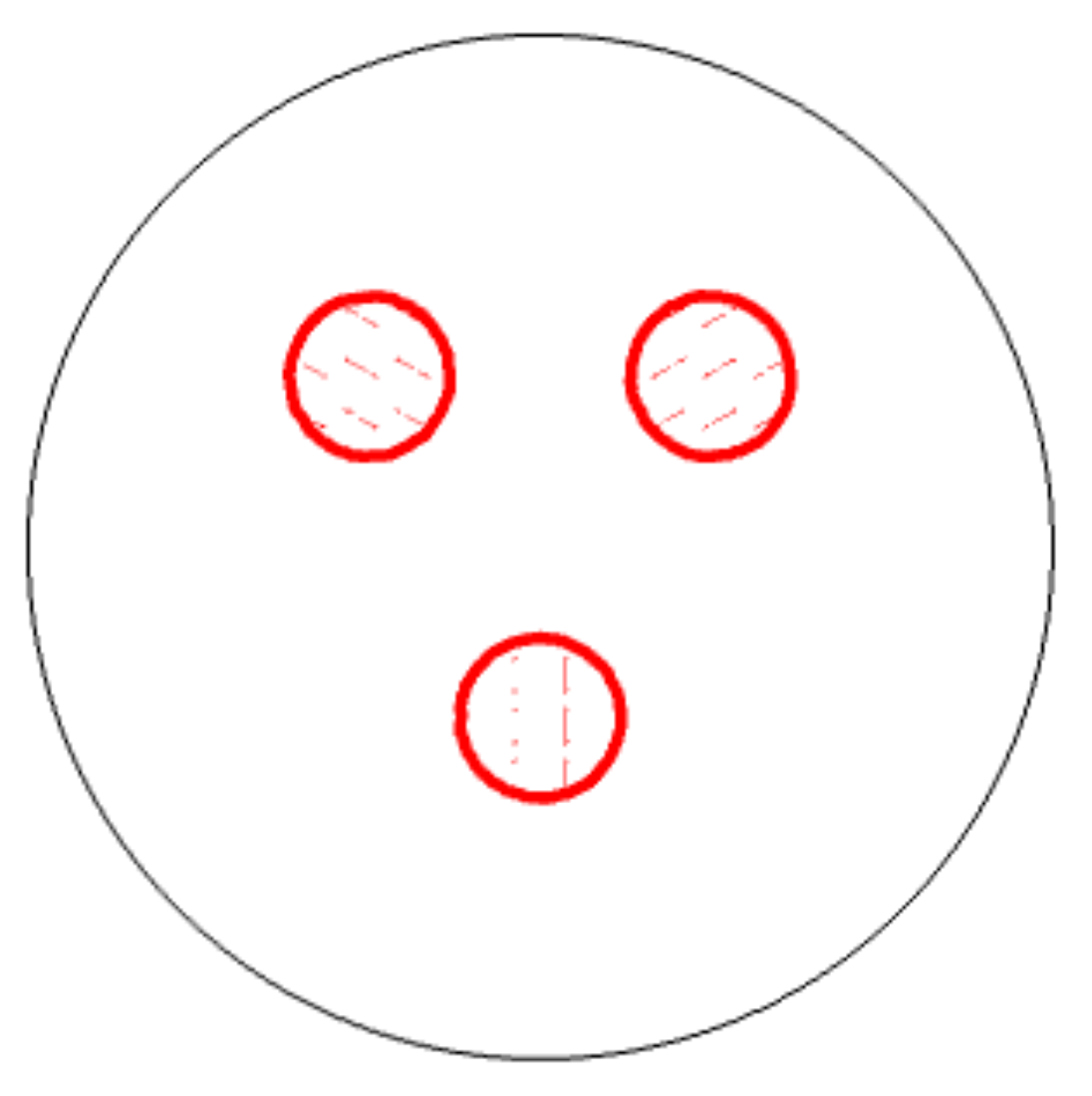}\quad
        \includegraphics[width=.4\linewidth,height=.4\linewidth]{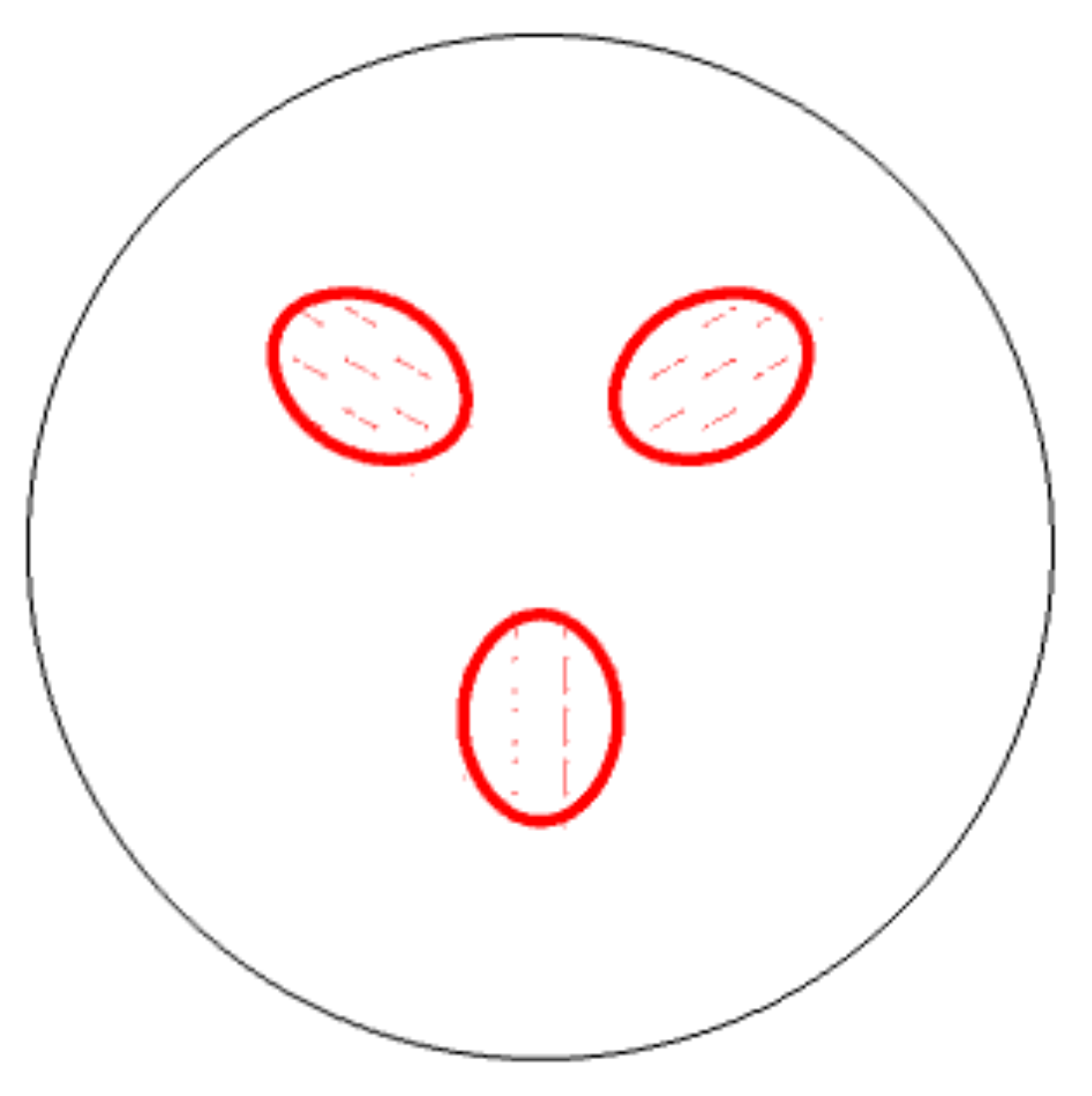}
        
        \vspace{3mm}
        
        \includegraphics[width=.4\linewidth,height=.4\linewidth]{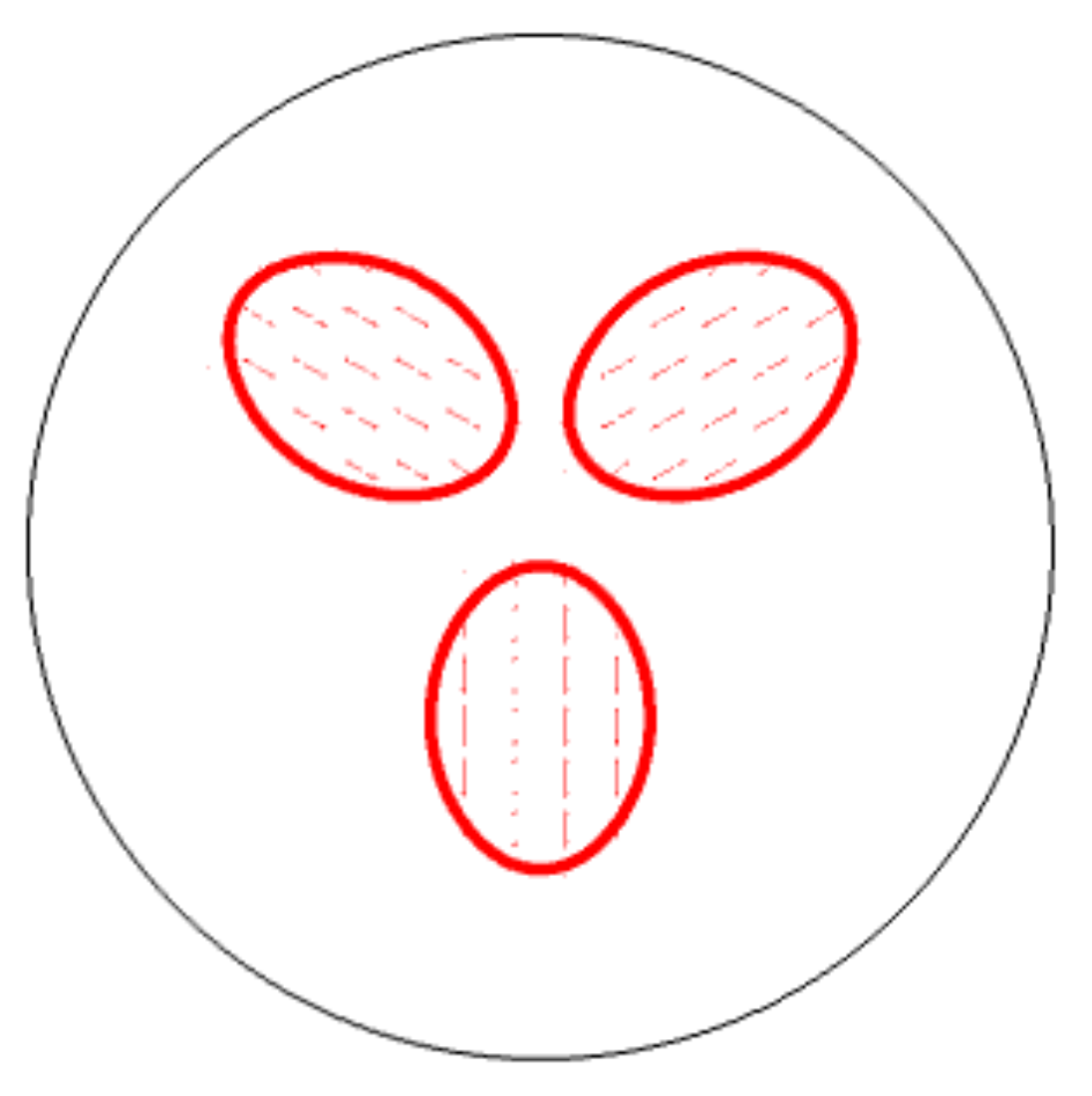}\quad \includegraphics[width=.4\linewidth,height=.4\linewidth]{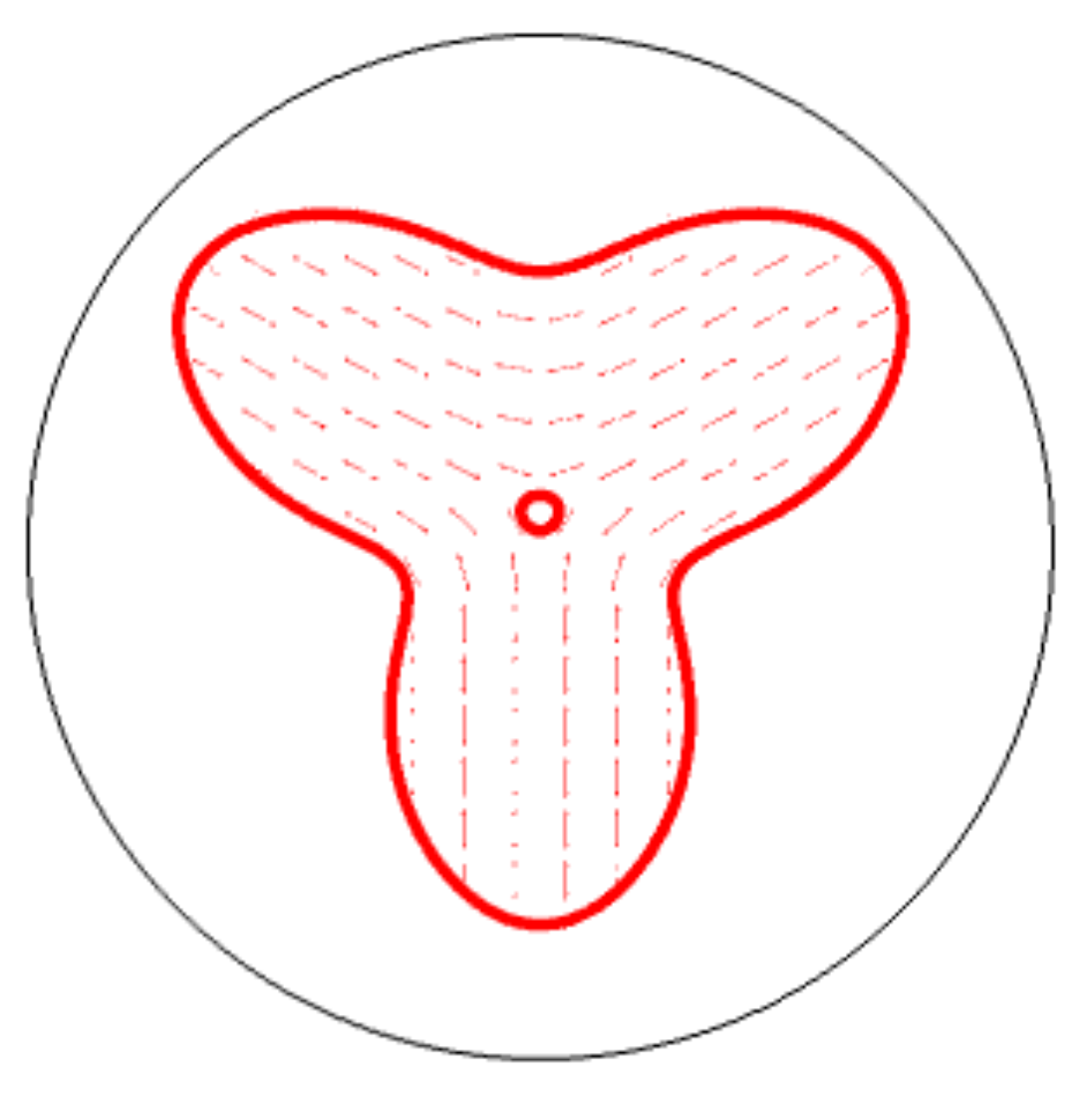}
        
        \vspace{3mm} 
        
        \includegraphics[width=.4\linewidth,height=.4\linewidth]{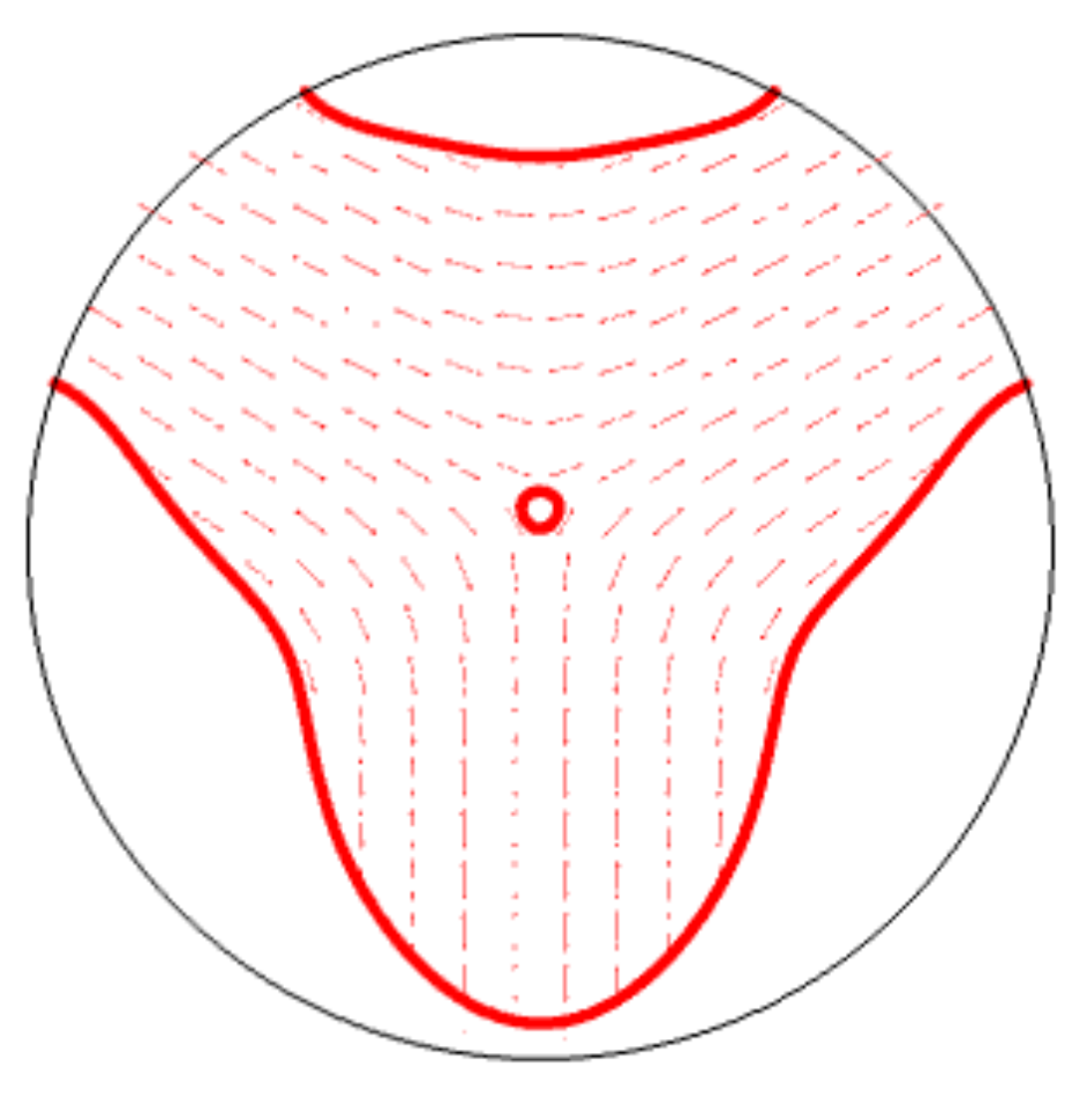}\quad \includegraphics[width=.4\linewidth,height=.4\linewidth]{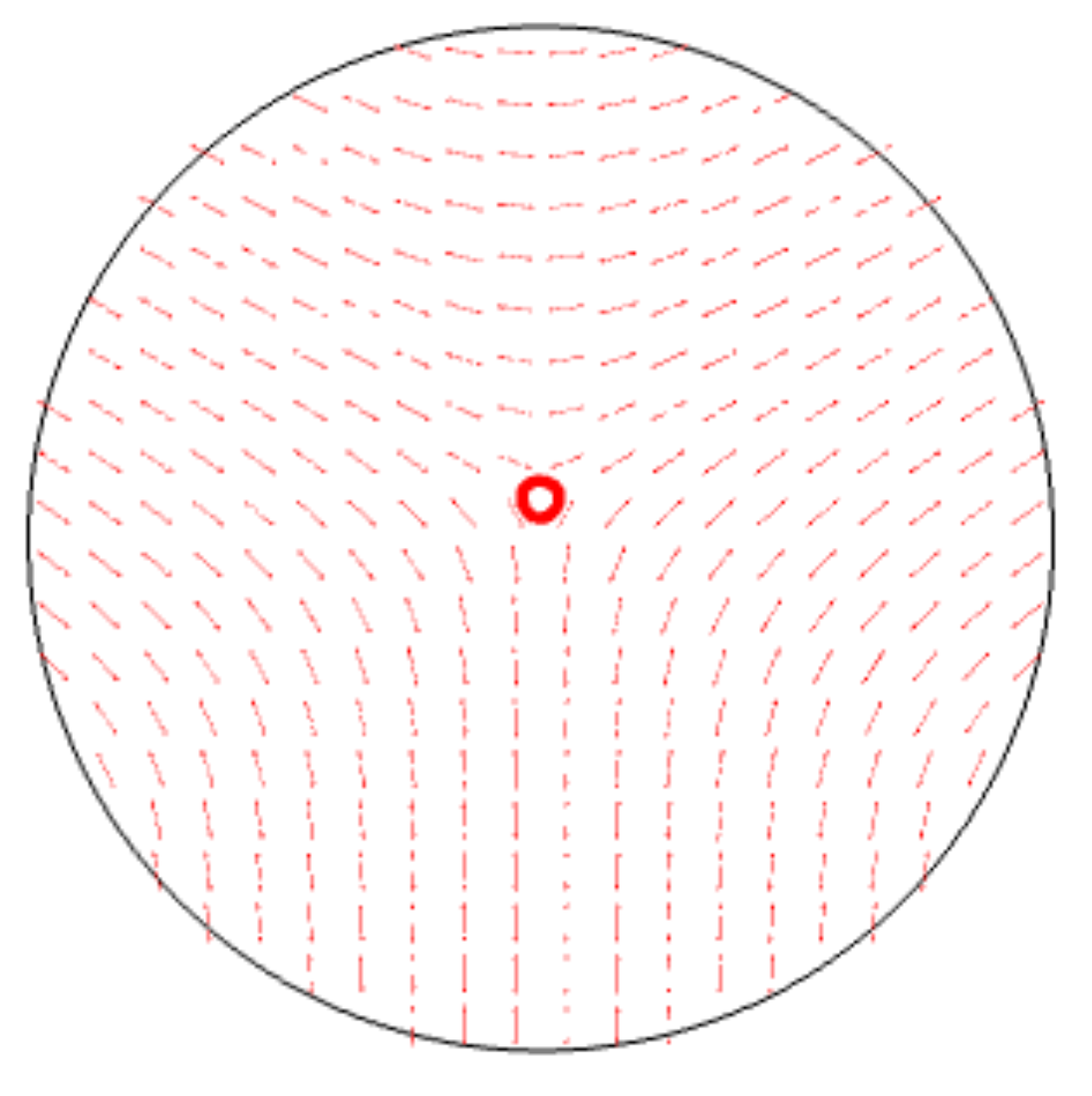}
        \caption{Simulated coalescence of three degree $0$ nematic tactoids. The thick red lines indicate the position of the interface.}
        \label{fig:someones_label}
    \end{figure}

    Finally, the Landau - de Gennes model can also be used to simulate the reverse situation when positive nematic tactoids nucleate in the isotropic phase, then grow and coalesce to form the nematic phase with embedded topological defects (cf. \cite{Kim_2013}). In Fig. \ref{fig:someones_label} the simulations were conducted subject to Neumann boundary data on $\partial\omega$ and assuming that $\omega$ has radius $1/2$, while $\alpha=0.01$. Three circular tactoids of different orientations were assumed to be present at the time $t=0$; in the course of the simulation, tactoids merged generating a single degree $-1/2$ defect. This situation closely resembles the original Kibble's model of strings formed in early universe through coalescence of domains with different ``phase" \cite{Kibble}.
    
\section{Acknowledgements}
DG acknowledges the support from DMREF NSF \\ DMS-1729538. OL acknowledges the support from DMREF NSF DMS-1729509. PS acknowledges the support from the Simons Collaboration grant 585520. 
\FloatBarrier 
\bibliographystyle{ieeetr}
\bibliography{references}
\end{document}